\documentclass[11pt,a4paper]{article}

\usepackage[utf8]{inputenc}

\usepackage{tikz,ifthen,xstring,calc,pgfkeys,pgfopts}
\usepackage{courier}

\usepackage{multirow}
\usepackage{array}
\usepackage{rotating}

\usepackage{hyperref}
\usepackage{url}

\usepackage{algorithm,algpseudocode}
\usepackage[linesnumbered,ruled,vlined,algo2e]{algorithm2e}
\usepackage{listings}

\usepackage{mathtools}
\usepackage{amsmath,amssymb,amsthm}
\usepackage{thmtools}

\usepackage{graphicx}
\usepackage{adjustbox}

\usepackage{todonotes}

\usepackage[charter]{mathdesign}
\usepackage[mathcal]{eucal}
\usepackage{bbm}

\usepackage{color, colortbl}
\definecolor{Gray}{gray}{0.9}
\definecolor{LightCyan}{rgb}{0.88,1,1}
\usepackage{subcaption}

\usepackage{setspace}
\usepackage[shortlabels]{enumitem}
\usepackage{scrextend}
\addtokomafont{labelinglabel}{\sffamily}

\usepackage[margin=1.2 in]{geometry}
\usepackage{pdflscape}

\usepackage[round]{natbib}
\usepackage{authblk}

\usepackage{titlesec}
\titleformat{\section}
	{\normalfont\Large\bfseries\filcenter}{\thesection.}{1 ex}{}
\titleformat{\subsection}%[runin]
	{\normalfont\normalsize\bfseries}{\thesubsection.}{1 ex}{}
\titleformat{\subsubsection}%[runin]
	{\normalfont\normalsize\bfseries}{\thesubsubsection.}{1 ex}{}
\declaretheorem[name=Theorem]{Thm}
\declaretheorem[within=section,name=Lemma]{Lem}
\declaretheorem[sibling=Lem,name=Definition]{Def}

\declaretheorem[sibling=Lem,name=Proposition]{Prop}

\declaretheorem[sibling=Lem,name=Corollary]{Cor}

\newcommand{\Real}{\operatorname{Re}}
\newcommand{\Imag}{\operatorname{Im}}

\newcommand{\Tr}{\operatorname{Tr}}

\newcommand{\Var}{\operatorname{Var}}

\newcommand{\Cov}{\operatorname{Cov}}

\newcommand{\Exp}{\operatorname{Exp}}

\newcommand{\Poi}{\operatorname{Poi}}
\newcommand{\Unif}{\operatorname{Unif}}

\newcommand{\calF}{\mathcal{F}}

\newcommand{\calN}{\mathcal{N}}

\newcommand{\CC}{\ensuremath{\mathbb{C}}}
\newcommand{\EE}{\ensuremath{\mathbb{E}}}

\newcommand{\NN}{\ensuremath{\mathbb{N}}}

\newcommand{\RR}{\ensuremath{\mathbb{R}}}

\makeatletter
\providecommand*{\diff}%
        {\@ifnextchar^{\DIfF}{\DIfF^{}}}
\def\DIfF^#1{%
        \mathop{\mathrm{\mathstrut d}}%
                \nolimits^{#1}\gobblespace
}
\def\gobblespace{%
        \futurelet\diffarg\opspace}
\def\opspace{%
        \let\DiffSpace\!%
        \ifx\diffarg(%
                \let\DiffSpace\relax
        \else
                \ifx\diffarg\[%
                        \let\DiffSpace\relax
                \else
                        \ifx\diffarg\{%
                                \let\DiffSpace\relax
                        \fi\fi\fi\DiffSpace}
\makeatother

\title{Periodic seismicity detection without declustering}

\author[1]{Timothy~Park}
\author[1]{Franz~J.~Kiraly}
\author[1]{Stephen~J.~Bourne}

\affil[1]{{\footnotesize Shell Global Solutions International B.V., 1031HW Amsterdam, The Netherlands}}

\date{\today}

\begin{document}
\maketitle
\doublespacing
\abstract{
Any periodic variations of earthquake occurrence rates in response to small, known, periodic stress variations provide important opportunities to learn about the earthquake nucleation process. Yet, reliable detection of earthquake periodicity is complicated by the presence of earthquake clustering due to aftershocks and foreshocks. Existing methods for detecting periodicity in an earthquake catalogue typically require the prior removal of these clustered events. Declustering is a highly uncertain process, so declustering methods are inherently non-unique. Incorrect declustering may remove some independent events, or fail to remove some aftershocks or foreshocks, or both. These two types of error could respectively lead to false negative or false positive reporting of periodic seismicity. To overcome these limitations, we propose a new method for detecting earthquake periodicity that does not require declustering. Our approach is to modify the existing Schuster Spectrum Test (SST) by adapting a test statistic for periodic seismicity to account for the presence of clustered earthquakes within the catalogue without requiring their identification and removal.

We describe the mathematical basis for this new test statistic, and the practical steps for applying this method to an earthquake catalogue. Using simulated earthquake catalogues we verify the expected performance of our Modified Schuster Spectrum Test (MSST) and compare it to the existing SST. In the absence of aftershocks, both methods yield similar true positive, false positive and false negative rates, as expected. In the presence of aftershocks, the existing SST false positive rates increase significantly whilst the MSST false positive rates remain essentially unaffected. We conclude by applying both methods to two observed earthquake catalogues with previously reported evidence for seasonal seismicity rates. The first is within the New Madrid Seismic Zone where the SST method finds statistically significant periodicities with and without declustering. Our MSST method finds no statistically significant evidence for any periodicity. This suggests that errors in declustering processes may confound periodicity tests that require aftershock and foreshock removal. Alternative tests that do not require declustering, such as the MSST, should enable a more reliable and sensitive means of investigating seismicity responses to periodic stress variations. The second observed catalogue is taken from the Himalayas and has been reported to show evidence of seasonal activity rates after declustering. We applied the SST and MSST to the data before declustering as well as the SST to the data after applying two different declustering techniques. We found that the evidence for seasonality when using the SST depended upon the parameter choices for declustering however with the MSST there is no ambiguity and it did not find statistically significant evidence for seasonality.
}

\clearpage

\section{Introduction}

Testing for periodicity in an earthquake catalogue is a common and important procedure in state-of-art seismological data analysis \cite{hernandez1999time} - for example, in the study of tidal/solar periodicities \cite{heaton1975tidal, tanaka2002evidence, cochran2004earth}, hydrospheric periodicities \cite{Ader2013, Johnson2017, Craig2017, JOHNSON2020}, or blast detection \cite{rydelek1994estimating}.
The earthquake periodicity testing problem is an instance of the general data scientific problem of testing for periodicity in an abstract series of events.
In this manuscript, we make two main contributions:
\begin{itemize}
\item Presenting what is, to our knowledge, the first seasonality test that can cope with earthquake clustering, which is also a formal hypothesis test with provable guarantees rather than a complete heuristic,
\item Validating its practical use by application on a selection of commonly known earthquake catalogues where seasonality is a question of interest.
\end{itemize}

To our knowledge, all state-of-art testing procedures with formal guarantees are subject to the implicit mathematical assumption of no aftershocks. Aftershocks are a phenomenon which is empirically well-validated, considered of practical importance, and scientifically well-studied since more than a century~\cite{utsu1995centenary}; in addition, the existence and severity of the issue was already pointed out, in 1897, in the original paper of Schuster~\cite[paragraph 5]{schuster1897lunar}. As a small thought experiment illustrates, actual presence of aftershocks where none are assumed can lead to both false detection and non-detection of periodicity: if aftershocks occur on a very short time scale compared to a yearly testing period, all shocks (incorrectly considered primary shocks) cluster in months and may create an illusion of seasonality; whereas if aftershocks occur on a time scale which is long compared to a daily testing period, periodic primary shocks may not be detected between aftershocks (incorrectly considered primary shocks) spreading more evenly across the day. More formally, in a periodicity testing procedure, the assumption of no aftershocks may lead to type I (false positives) and type II (false negatives) errors, caused by null hypothesis mismatch.

Current common techniques for detecting earthquake periodicity all rely on core testing procedures which assume that there is no earthquake clustering, such as based on the Schuster test \cite{schuster1897lunar,Ader2013}, after prior heuristic removal of clustered events from the catalogue ensuring applicability of the core testing procedures. Here, ``heuristic'' means that these prior procedures come with no guarantees of correctness, e.g., that no strong bias is introduced. Hence, the state-of-art suffers from \emph{severe limitations}, as declustering is an imprecise procedure that relies on judgement-based choices \citep{Molchan1992a,Zaliapin2008} that typically vary between applications, likely bias the periodic signals that are sought, and weaken any scientific conclusions based on them to the strength of the weakest link in the argumentation chain which is typically, declustering choices. The method of \cite{Dutilleul2015} is one example which can be applied to data without declustering however this is a heuristic method and is applied to monthly event counts rather than event times.

Our proposed testing procedure, MSST, succeeds in removing the strong assumption of no clustering from the core testing procedure, in favour of a much weaker assumption of no secondary or higher-order aftershocks. As secondary aftershocks typically account only for a very small fraction of shocks, or can be removed much more easily, this is much less of a restriction than the far stronger assumption of no clustering, and much less of a problem than heuristic testing procedures without any proper mathematical guarantees. Therefore our testing procedure is significantly less susceptible to errors in periodicity detection due to clustering than existing state-of-the art procedures.
Furthermore, we argue that the assumption of no secondary aftershocks is not a restriction in-principle, as it can also be treated in our mathematical framework - though at the cost of substantial mathematical overhead in this manuscript which we chose to avoid in favour of readability and clarity of exposition. We leave this for future work. We discuss the two contributions in more detail below.

\subsection{Data scientific contribution: MSST}

We discuss our methodological contribution in the context of previous methodological advances. 
Historically, abstract hypothesis tests for periodicity in event sequences are closely linked with the seismology application.
The most common approach in seismology applies frequentist hypothesis tests to the event sequence or a derived time series, see~\cite{hernandez1999time} for an overview. To our knowledge, this is also the only kind of framework to which formal guarantees have been derived, \emph{i.e.}, guarantees on type I and type II errors, or guarantees on quantifiers of evidence arising from the testing procedure.

Most variants rely on procedures closely related classical work of Schuster and Fisher~\cite{schuster1897lunar,fisher1929tests}. Differentiation occurs through the specific approximation of the statistic, and whether testing is for a specific, pre-determined period, or multiple periods at once.
The methodological backdrop is somewhat subtle - there are multiple non-identical procedures which are called ``the Schuster test'' in literature; and the procedure known under this name today was first introduced by Heaton~\cite{heaton1975tidal} in 1975 and is not identical with the original procedure by Schuster and Fisher.
Further, to our knowledge, no formal proof for properties of this latter Schuster test exist, though they are not difficult to adapt from the original work of Schuster and Fisher - which we carry out as a subsidiary contribution in this manuscript.

We provide a more technical discussion of the data scientific literature in Section~\ref{sec:prev}, as making precise the differences in methods and discussing the trail of ideas requires introduction of some mathematical notation.
For our main contribution, it is important to note that all testing procedures discussed above assume that there is no clustering, i.e., the methodological framework of an inhomogeneous Poisson process. Treatment of clustering, in literature, is predominantly heuristic, e.g., reliant on windowing or bootstrap simulation procedures without any formal contraposition, null hypothesis, or type I/II guarantees.

Our proposed method, MSST, is to our knowledge the first method to provide such guarantees in the case of a general process with aftershocks. It is based on theoretical results on asymptotic properties of a first-generation inhomogenous Poisson process with second-generation aftershocks. The null hypothesis is that the first-generation process is homogeneous, and type I/II guarantees (more precisely: control of type I error given fixed type II error) is obtained in the common frequentist testing framework of asymptotic approximation. This contains an additional semi-heuristic step of assuming sufficient regularity in the Schuster spectrum.
Technical details and a summary of assumptions are presented in Section \ref{sec:newTest}.

\subsection{Geoscientific contribution: MSST}

Recent advances in earthquake data processing using methods such as template matching \citep{Yoon2015,Li2018} and convolutional neural networks \citep{Perol2018,Ross2019} mean the quality and size of earthquakes catalogues are growing rapidly. This creates new opportunities to detect the amplitude and phase of small periodic variations in earthquake nucleation rates driven by cyclic stresses within the earth's crust. Such observations are important as constraints on frictional fault failure processes \citep{Beeler2003a,Ader2014}, as indicators of mean stress changes \cite{Tanaka2004,Ogata2005,Ogata1999}, and as indicators of proximity to failure in a future earthquake \cite{Tanaka2012}.

There are multiple sources of cyclic stress that load seismogenic faults: hydrospheric, atmospheric, thermal, tidal, solid Earth tides \citep[e.g.][]{Johnson2017,Johnson2017a,Heki2003, Tolstoy2002,Metivier2009}, and anthropogenic activities such as seasonal variations in water reservoir levels \citep{Simpson1988,Talwani1997} and natural gas extraction, such as the Groningen field \citep{Bourne2014a,Bourne2017a,Bourne2018}.
For semidiurnal periods, micro-earthquakes regions of seafloor volcanic and hydro-thermal activity do correlate with ocean tides \citep{Wilcox2001,Tolstoy2002,Stroup2007}, as do shallow tectonic earthquakes \citep{Tanaka2002,Wilcock2009}.
However the evidence for seismicity triggered by semidiurnal solid Earth tides is mixed; with reports of weak to no correlations \citep{Shudde1977,Heaton1982,Vidale1998,Beeler2003a}, strong positive correlations for all focal mechanisms \citep{Metivier2009} or only thrust-fault mechanisms \citep{Cochran2004}, or only normal-fault mechanisms \citep{Tsuruoka1995}. Tanaka reports significant triggering of earthquakes by semi-diurnal solid Earth tides prior to the 2011 $M_w$ 9.1 Tohoku-Oki earthquake \citep{Tanaka2012}, and a systematic increase in the amplitude of semidiurnal seismicity with proximity to failure in both time and space over the preceding 10~years and out to epicentral distances of 100~km. \cite{Wang2015} repeats this study but finds no clear evidence for any semi-diurnal or semi-monthly earthquake periodicity. The key difference between these contradictory results are judgement-based choices about declustering. As the characteristic time of earthquake clustering overlaps strongly with these tidal periods it is always possible that declustering may inadvertently bias the result one way or the other leaving the outcome uncertain.

For seasonal periods, evidence of seismicity triggered by hydrologically-driven seasonal stress cycles are reported in Japan \citep{Heki2003,Ueda2019}, the Himalayas \citep{Bollinger2007,Bettinelli2008,Ader2013}, the New Madrid intra-plate region of North America \citep{Craig2017}, California \citep{Johnson2017}, the volcanic centres of western US \citep{Christiansen2005}, the Balkans \citep{Muco1999}, India \citep{Smirnov2018}, and icequakes in Switzerland \citep{DalbanCanassy2016}.
This apparent stronger correlation of seismicity with seasonal loads rather diurnal solid Earth tides is attributed the earthquake nucleation time-scale being much longer than the diurnal time-scale \citep{Beeler2003a}, but smaller than the seasonal time-scale. All these results however share the same limitation by depending on uncertain declustering choices that may leave a residual bias that appears as apparent seasonal seismicity \cite{Ader2013}. The periodicity of acoustic emissions in laboratory creep experiments do show a similar transition under cyclic loads \citep{Chanard2019}, but the scaling of this result to geological faults is uncertain.

Methods for detecting earthquake periodicity without declustering would be very useful to reduce bias and to improve both temporal and spatial resolution of periodic seismicity responses to stress changes. In this paper we present what we believe are the first theoretical results on the Schuster test statistic in the presence of aftershocks. We present some experiments on synthetic data and a real earthquake catalogue, corroborating that the impact of omitting the important assumption is not only practically meaningful, but potentially very serious. We also discuss how to diagnose the severity of the missing aftershock assumption for a given seismological use case, as well as possible pathways (subject to potential alternative assumptions) for adapting the Schuster test to real world earthquake catalogues with aftershocks. Technical details and assumptions are described in Section~\ref{sec:newTest}.

\subsection{Structure of this manuscript}

The manuscript is structured as follows:
Section \ref{sec:prev} gives a brief description of the existing Schuster spectrum algorithm. In Section \ref{sec:newTest} we present a theoretical description of the effect of aftershocks on the test statistic for the Schuster spectrum test. We then describe a method for correcting the test to account for this effect by making use a a spline based fit to estimate the unknown aftershock effects. Finally in Section \ref{sec:results} we show the results of applying our modified test to a range of simulated earthquake catalogues and to two observed example of seasonal seismicity due to hydrospheric loading from the New Madrid region of North America \citep{Craig2017} and the Himalayas \citep{Ader2013}.

\section{Prior Work on the Schuster test}\label{sec:prev}

\subsection{The Schuster test - naming ambiguities and mathematical summary}
\label{sec:lit.mathex}

In literature, somewhat inconsistent use of terminology surrounds ``the Schuster test'', as in referring to procedures that are closely related mathematically and algorithmically, but not exactly identical. While this has constituted no severe issue in the relevant applications, it makes a precise literature discussion difficult as multiple terms are used for multiple non-identical procedures, interchangeably. As the latter issue arises from somewhat subtle, mathematical points, we start by briefly introducing some necessary mathematics.\\

The procedure which we in this manuscript refers to as ``the Schuster test'' is a test for periodicity in an \emph{event sequence}, i.e., a series of observed event times $T_1,\dots, T_N$ (presented in uniformly random order). Periodicity at a radian frequency $\omega$ is tested through an effect size called ``squared Schuster distance'', defined as
$$D^2(\omega):= |D(\omega)|^2, \;\mbox{where}\quad D(\omega) := \sum_{i=1}^N \exp(\imath \omega T_i)$$
is the ``Schuster distance''. Under the null hypothesis of no periodicity, $\exp(\imath \omega T_i)$ is uniformly distributed on the complex unit circle, and it can be shown that $D^2(\omega)$ is approximately $\Exp(N)$-distributed, giving rise to a p-value-like test statistic, testing against the alternative of periodicity where $D^2(\omega)$ is larger. A detailed discussion with explicit proofs and derivations can be found in Appendix~\ref{sec:app-schuster}. There may be small variations in how the final p-value is computed, e.g., through a choice of asymptotic approximation, therefore, formally, there are multiple possible variants of this ``Schuster test''.

A second, closely related method is called the ``Schuster/Fisher procedure'', or the ``Fisher g-test''. This is a test of periodicity in a \emph{time series}, i.e., a series of observed real values $X_1,\dots, X_N$, at pre-determined times $t_1,\dots, t_N\in \RR$, which are equally spaced, i.e., all the differences $t_{i-1}- t_i, i=1\dots N-1$ are equal. The effect size is the Fourier spectrum energy at radian frequency $\omega$, that is.
$$E(\omega) := \left|\sum_{i=1}^N X_i\exp(\imath \omega t_i)\right|^2.$$
Under the null hypothesis of no periodicity, it can be shown that $E(\omega)$ is approximately exponential with known mean, giving rise to a suitable test statistic. In addition, if the $X_i$ are normally distributed, an exact statistic is available - ``Fisher's g'' (which otherwise only yields a good approximation).

As time series are not the same as event series, the two testing procedures are not identical - however, they are closely related. If, starting at the symbols in ``the Schuster test'', the times $T_i$ are all replaced by the closest (``rounded'') $t_j$, and $X_j$ is defined as the number of $T_i$ that are replaced by $t_j$, the Schuster distance of the rounded event series, and the Fourier spectrum energy of the time series of counts $X_j$ observed at $t_j$, are mathematically identical. Conversely, starting with symbols in ``the Schuster/Fisher procedure'', if all the $X_j$ are integers, one can produce an event series where $t_j$ appears $X_j$ times, and the same identification holds in reverse. Thus, the ``event series'' setting can be interpreted as the limit of infinitesimally small bins, arising from a time series of bin counts.
It is also interesting to note that, under these identifications, the mathematical argumentation, derivations of test statistics and p-values and their properties, largely coincide.

In below discussion, we will refer to the Schuster test or Fisher g-test procedure (no quotes) when referring to the mathematical class of procedures as discussed above. We will refer to ``the Schuster test'', ``Schuster/Fisher procedure'', or ``Fisher g-test'' when referring to the terms, not the procedures, sparingly and only insofar it is necessary for the discussion.

\subsection{Evolution of ``the Schuster test(s)'' and related procedures}
\label{sec:lit.schusterschusterschuster}

% about history of the test itself

The Schuster test (as defined in Section~\ref{sec:lit.mathex} or in Appendix~\ref{sec:app-schuster}), seems to first appear as an original idea in~\cite{heaton1975tidal} as a heuristic algorithmic procedure without proof, citation of earlier sources, or mention of Schuster, or Fisher, or any Schuster/Fisher related terminology.
The seminal papers of Schuster~\cite{schuster1897lunar} and Fisher~\cite{fisher1929tests}, frequently cited as original sources for ``the Schuster test'', introduce in fact not the Schuster test, the test for periodicity in event sequences, but the Fisher g-test, that is, the test for periodicity in time series. Fisher~\cite{fisher1929tests} greatly extends Schuster's original work~\cite{schuster1897lunar}, formulated for time series of counts (arising from earthquakes).
Confusingly, despite the Schuster test not appearing in the original work of Schuster and Fisher, or any other work of Schuster, most contemporary sources refer to the Schuster test indeed as ``the Schuster test'' (using Schuster's name, ``Schuster'', as an epithet).
While the Fisher g-test has also found applications in many domains outside geoscience, use of the Schuster test is mostly localized to geoscience - although, confusingly, both have been historically, and recently, applied to important applications in seismology.

As regards formal mathematical proofs, \emph{e.g.}, on the distribution of the test statistic and its asymptotics, Schuster's and Fisher's original work~\cite{schuster1897lunar,fisher1929tests}, as well as subsequent theoretical work, seems to focus entirely on equally spaced time series, i.e., the Fisher g-test. In contrast, the Schuster test appears in literature mostly as an algorithm and heuristic, with literature frequently (and incorrectly) citing Schuster's and Fisher's work (on a different test, the Fisher g-test) as a theoretical justification.
To the best of our knowledge, our Appendix~\ref{sec:app-schuster} is the first complete formal treatment of the Schuster test, its properties, and the first proper justification for its use in its event series setting - where, of course, the argumentation is, as expected, largely but not entirely congruent with the parallel results for the Fisher g-test.

One additional noteworthy area of research and results is the treatment of the multiple testing problem, arising from testing for ``some periodicity'' (across many radian frequencies $\omega$) as opposed to ``on specific periodicity''. While this is now relatively classical for the Fisher g-test, see for example \cite{Brockwell2013}, developments for the Schuster test are relatively recent, with prevalent miss use in the geoscientific domain, as for example carefully highlighted by Ader and Avouac~\cite{Ader2013}. In the same article~\cite{Ader2013}, a multiple testing correction procedure is suggested for the special case of sinusoidal variations, but this is not generally valid for all periodic variations.

\section{The Schuster Test Statistic in the Presence of Aftershocks}\label{sec:newTest}

In this section we describe our proposed aftershock correction for the Schuster test, based on derivations in the case of a relatively general aftershock model. We start in Section \ref{sec:genProc} with the definition of such a process which will be the basis of our theoretical results. In Section \ref{sec:distSpec} we present some theoretical results based on this process. These results demonstrate analytically why the standard SST is not suited to the case where there are significant aftershocks present in the earthquake catalogue.

\subsection{Earthquake catalogue generative process}
\label{sec:genProc}

For the purpose of this study, we adopt the following mathematical model as an aftershock process:
\begin{itemize}
\item We observe shocks at times $T_1,\dots, T_N$, all at some time in $[0,\infty)$. There is at most one shock at a given time.
\item All shocks are either one of $M$ primary shocks, or one of $A_i$ aftershocks of the $i$-th primary shock. In particular, this implies that $N = M + A_1 + \dots A_M$.
\item The precise labelling of observed shocks, i.e., whether a shock is a primary shock or an aftershock of the $i$-th primary shock, is a-priori unknown to the observer.
\item Aftershocks are always an aftershock of one and only one associated primary shock. Aftershocks themselves have no aftershocks.
\item There is a fixed observation window $[0,1]$ for the catalogue (which can be re-scaled or shifted), in which all the primary shocks are observed. Aftershocks, but not primary shocks, may lie in $[1,\infty)$. This is a plausible assumption if the typical aftershock occurs much closer to their primary shocks than the observation window is wide. We will discuss and remove this simplifying assumption later on.
\item The primary shocks are assumed to be sampled from a (possibly heterogeneous) Poisson process, henceforth referred to as the \emph{primary shock process}, assumed to have some intensity functional $\lambda_0$ (vanishing outside the observation window $[0,1]$).
\item All aftershocks associated to one given primary shock are assumed to be sampled from a (possibly heterogeneous) Poisson process, henceforth referred to as the \emph{aftershock processes}.
\item The statistical dependency between aftershocks is mediated entirely through their respective primary shocks; that is, conditional on primary shocks (being known/observed), different aftershock processes are statistically independent.
\item Aftershock processes are assumed causal, i.e., have zero intensity at time before their associated primary shock (that is, the intensity functional $\lambda$ for the aftershock process of a primary shock $T_i$ satisfies $\lambda(t) = 0$ for any $t < T_i$).
\item Aftershock processes are assumed identical in distribution and intensity, on a shifted time axis which has the associated primary shock at time zero (that is, the intensity functional for the $i$-th aftershock process, of a primary shock at time $T_i$, is assumed to be $t\mapsto \lambda(t-T_i)$ for some functional $\lambda$, where the dependency on the primary shock is through $T_i$ and through $T_i$ only).
\end{itemize}

A mathematically formal exposition of this setting is found in Appendix~\ref{sec:app-proof.setting}. Formally, the process defined through the assumptions above is equivalent to a Hawkes process where aftershocks cannot have aftershocks. Hence our setting constitutes a simplification, but it is sufficient to highlight some key phenomena in the presence of aftershocks.

\subsection{Notation conventions}
The complex unit will be denoted by the symbol $\imath$.
Exponential distributions will be denoted as parametrised by inverse rate or decay parameter, \emph{i.e.}, $\Exp (\lambda)$ denotes the exponential distribution with expectation $\lambda$. 

\subsection{The Schuster distance in the presence of aftershocks}
\label{sec:distSpec}

We continue analysing, as a well-known test statistic, the squared Schuster distance. This is defined as $D^2(\omega):= |D(\omega)|^2$, where $\omega$ is some radian frequency, and $D(\omega)$ is the Schuster distance, defined as
$D(\omega) := \sum_{i=1}^N \exp(\imath \omega T_i).$
Often, $D$ is also seen parametrised by period length $k:=\frac{2\pi}{\omega}$ instead of frequency $\omega$.

Under the common null hypothesis of no seasonality overall, the primary intensity functional $\lambda_0$ is constant, i.e., $\lambda_0(t) = \nu_0$ for any $t\in[0,1]$ and some $\nu_0\in [0,\infty)$. For seasonality at a specific radian frequency $\omega$, the nulls is weaker - in this case, it is assumed only that $D(\omega)$ behaves as if $\lambda_0$ were constant, i.e., any angle $\exp(\imath \omega T_i)$ is uniform on the complex unit circle. Under this null hypothesis, the following holds:

\begin{Prop}\label{prop:exd}
Assume that the primary aftershock process (as specified above) has no periodicity at frequency $\omega$, that is, $\exp(\imath \omega T_i)\sim \Unif \{x\in \CC\;:\;|x|=1\}$, for all $i$.\\
As above, denote by $\lambda$ the aftershock intensity functional, and by $\nu$ is the total rate constant of the aftershock process, i.e., $\nu =\int_0^\infty \lambda(t)\diff t.$\\
Denote by $\chi$ the characteristic function to the pdf $\nu^{-1}\lambda$; equivalently,
$$\chi(t):=\nu^{-1} (\calF \lambda)\left(-\frac{t}{2\pi}\right),$$
where $\calF\lambda$ is the Fourier transform (unitary normalization convention) of the aftershock intensity functional $\lambda$. Then:
\begin{enumerate}
\itemsep-0.2em
\item[(i.a)] it holds for the expected Schuster distance that $\EE[D(\omega)] = 0$
\item[(i.b)] it holds for the variance of the absolute Schuster distance that $\Var[|D(\omega)|] = \Var[D(\omega)] = \EE[D^2(\omega)]$
\item[(ii)] it holds for the expected squared Schuster distance spectrum that
\begin{align}\label{eq:prop1}
	\EE[D^2(\omega)] &= \nu_0\cdot\nu + \nu_0\cdot\left|1+ \nu\cdot \chi\left(\omega\right)\right|^2.
\end{align}
\item[(iii)] if the numbers $M$ of primary shocks and $A_1,\dots, A_M$ of aftershocks are considered fixed, the expected squared Schuster distance (conditional on observed numbers) can be expressed as
\begin{align}
	\EE[D^2(\omega)|M,A_1,\dots, A_M] = (N-M) \left(1 - \left|\chi\left(\omega\right)\right|^2\right) + \sum_{m=1}^{M} \left|1 + A_m\cdot \chi\left(\omega\right)\right|^2.
\end{align}
\end{enumerate}
\end{Prop}

A proof can be found in Appendix Section~\ref{sec:app-proof.Schuster_expectation}, where this Proposition~\ref{prop:exd} appears as part of Proposition~\ref{prop:exd-appx}.

The term $\left(1 - \left|\chi\left(\omega\right)\right|^2\right)$ is also closely related to the variance of the aftershock process:

\begin{Lem}\label{Lem:aftvar}
Let $Z$ be a random variable distributed according to the pdf $\nu^{-1}\lambda,$ as considered in Proposition~\ref{prop:exd}. Then the following hold:
\begin{enumerate}
\itemsep-0.2em
\item[(i)] $\Var (\exp(\imath \omega Z)) = 1 - \left|\chi\left(\omega\right)\right|^2$
\item[(ii)] $\Var (\exp(\imath \omega Z)) \le \omega^2 \Var(Z).$

\end{enumerate}
\end{Lem}
\begin{proof}
(i) is a consequence of Lemma~\ref{Lem:Fourier}~(i) in the Appendix; (ii) is a consequence of Proposition~\ref{Prop:phasor}~(v) in the Appendix.
\end{proof}

(for convenience, variance of a complex random variable, as used in Lemma~\ref{Lem:aftvar}~(i), is introduced and discussed in Appendix~\ref{sec:complexRV})

\subsection{Empirical use and estimation}
\label{sec:estimation}

Without further methodology, Proposition~\ref{prop:exd} is not practically usable: the objects $\nu_0$, $\nu$, $\chi$, $\calF\lambda$ are unknown to the observer, and are in general not straightforward to estimate without knowledge of which shocks are primary or aftershocks. Before proceeding with how the above is usable for a correction to the Schuster test, we would like to point out how the quantities are physically interpretable and obtainable from low-assumption plug-in estimates:
\begin{enumerate}
\itemsep-0.2em
\item[(i)] $\nu_{0}$ is the expected number of primary shocks, and $\nu\nu_0$ is the expected number of aftershocks. As the observed number of shocks, $N$, is a realisation of the sum $\nu(1+\nu_0)$, plug-in estimates $\widehat{\nu}_0:= \gamma N$ and $\widehat{\nu} := (\gamma^{-1}-1)$ may be obtained for any assumed or estimated fraction of primary shocks $\gamma$.
\item[(ii)] It holds that $\nu\cdot \chi(\omega)= (\calF\lambda)\left(k^{-1}\right)$, the Fourier transform of $\lambda$. The latter describes phase and energy of the aftershock process (kernel). This may be directly obtained from an assumed or estimated explicit form of the aftershock process (kernel). It is closely related to variance of the aftershocks (and aftershock phasors) as discussed in Lemma~\ref{Lem:aftvar}, also see Lemma~\ref{Lem:Fourier} and Proposition~\ref{Prop:phasor} in the appendix. If it is known which shocks are aftershocks of the same shock, this term can thus also be obtained from sample variance estimation.
\end{enumerate}

\subsection{Example: exponential aftershock model}
We illustrate the impact of aftershocks on the original Schuster procedure, and thus possible corrections to it, by considering a simple example in the context of Proposition \ref{prop:exd}. Within observational seismology, temporal clustering of aftershocks was described by the long-standing Omori law \cite{Omori1894}, $\lambda(t) = K(t+c)^{-1}$, where $K$ and $c$ are constants inferred from earthquake catalogues. More recently this was generalized by the Modified Omori law \cite{Utsu1995},  $\lambda(t) = K(t+c)^{-p}$. The constant $c$ is typically smaller than 1 day, and the constant $p$ varies in the range 0.9–1.9. This inverse-law parameterization of the aftershock model does not lead to simple algebraic expressions within this analysis.
For simplicity, we will illustrate our method with exponentially-distributed aftershocks as an initial approximate representation of the Modified Omori law. We anticipate these results may be readily extended to include the Modified Omori law as its Fourier transform also has a closed-form expression, albeit somewhat more complex.

Consider the aftershock model with an exponential form of $\lambda(t)$:

\begin{align}\label{eq:afm}
	\mbox{Let}\; \lambda(t) &= \frac{A}{\tau} \cdot e^{-t/\tau}\cdot \mathbbm{1}[t\ge 0],\;\mbox{with parameters}\;\tau, A\in \RR_+.\; \mbox{Then,} \nonumber\\
	(\calF \lambda)(\omega) &= \frac{A}{\omega \tau^2  +\imath},\quad
    \nu = \int_0^\infty \lambda(t) \diff t = A, \nonumber\\
    \EE [D^2(\omega)] &= \nu_0\cdot A + \nu_0\cdot \left|1 + \frac{A}{\omega \tau  +\imath}\right|^2, \nonumber\\
    %\left|1+(\calF \lambda)(\omega)\right|^2 = \frac{A^2+(A+\omega \tau)^2 }{1+\omega^2 \tau^2}
    \EE [D^2(\omega)|M,A_1,\dots, A_M] &= \frac{N-M }{1+(\omega\tau)^{-2}} + \sum_{i=1}^M \left|1+ \frac{A_m}{\omega \tau  +\imath}\right|^2
\end{align}
For illustration of the impact of aftershocks, we consider some corner cases:

\paragraph{No Aftershocks:} In the case where the catalogue does not contain any aftershock events, we should recover well-known properties of the SST procedure. Mathematically, no aftershocks means $A=0$, and $A_i=0$. In this case $\mbox{E}[D^2(\omega)] = \nu_0$, which is the main shock rate constant, with the natural plug-in estimate in this case being $N = \mbox{E}[D^2(\omega)|M]$, the total number of earthquakes observed. This is the expectation used in the SST procedure, as expected.

\paragraph{Long characteristic aftershock duration:} This should intuitively agree with a situation of ``maximal decoupling'', i.e., for all practical purposes aftershocks are unrelated to the primary shocks. Mathematically, if the characteristic duration (also known as the intensity decay constant) of aftershocks, $\tau$, is very long as compared to the period, $k$, the intensity function $\lambda$ will become increasingly flat. In the limit $\tau\rightarrow \infty$, the Fourier transform, $\calF\lambda$ will approach a delta function supported at 0, for finite values of $k$. In equation \eqref{eq:prop1} the limit of the squared Schuster distance becomes
\begin{align}
	\lim_{\tau\rightarrow\infty}\mbox{E}[D^2(\omega)] = \nu_0\cdot A + \nu_0\cdot 1 = \nu_0\cdot (\nu + 1)
\end{align}

which is simply the expected number of shocks, overall.\\
As one may intuitively expect, one recovers the case of no aftershocks when considering aftershocks as primary shocks (as one should, since they are in fact completely decoupled). Again, $N$ is the natural plug-in estimate.

\paragraph{Short aftershock duration:} This should intuitively agree with ``counting the primary shock with multiplicity'', as aftershocks are infinitesimally instantaneous and virtually indistinguishable from the primary shock. Mathematically, if the characteristic duration of aftershocks, $\tau$, is very short as compared to the period, $k$, the intensity function $\lambda(t)$ approaches $A$ times a delta function. In this case the Fourier transform $\calF\lambda$ will approach a constant function with magnitude of $A$. More formally, the limit of equation \eqref{eq:prop1} can be shown to be equal to
\begin{align}
	\lim_{\tau\rightarrow 0} \mbox{E}[D^2(\omega)|M,A_1,\dots, A_M] = \sum_{i=1}^M (1+A_m)^2,
\end{align}
which corresponds to the case without aftershocks, but counting each primary shock with weight $(1+A_m)$ accumulated from counting all its aftershocks as part of the primary shock.

Outside such corner cases, that is, for a general choice of $\lambda(t),$ it is not easy to simplify equation \eqref{eq:prop1}, and therefore the expected value of $D^2(\omega)$ will vary with frequency. Though the important point to note is that the transition behaviour between the (long and short aftershock period) corner cases is continuous, as Equations~\ref{eq:afm} quantitatively confirm. That is, as the aftershock intensity becomes more peaked, the effective multiplicity with which primary shocks count grows; whereas, as aftershock intensity becomes flatter, the effective number of primary shocks increases.

If the actual primary shocks are correctly identified as such in the theoretical model, approaching the former extreme (``short aftershock duration'') causes type I errors, i.e., detecting periodicity where there is none. Qualitatively, approaching the latter extreme (``long aftershock duration'') causes type II errors, i.e., not detecting periodicity where there is.

Interestingly, the ``short aftershock duration'' corner case was also already discussed in the original manuscript of Schuster~\cite[paragraph 5]{schuster1897lunar} (for the closely related Schuster/Fisher procedure, see Section~\ref{sec:lit.schusterschusterschuster}) - as a criticism to, or response to, coefficients derived by Knott~\cite{knott1888earthquakes} about a decade earlier, alleging what in modern terminology are type I errors.

\subsection{Modification to the Schuster Spectrum Test}
\label{sec:adaptedTest}

An aftershock corrected Schuster test may be derived from comparing the (large sample) distribution of the Schuster spectrum in the presence and absence of aftershocks:

\begin{Prop}
\label{Prop:aftershockschuster}
Assume the null hypothesis of no periodicity holds, that is, the primary shock process is a homogeneous Poisson process (the intensity functional $\lambda_0$ is constant). Then, the following approximations hold:
\begin{enumerate}
\itemsep-0.2em
\item[(i)] if there are no aftershocks, then $\frac{1}{N}D^2(\omega)\approx \Exp\left(1\right)$.
\item[(ii)] in the general case, $\frac{1}{N}D^2(\omega)\approx \Exp\left(1-\kappa^{-1}+ \kappa^{-1}\left| 1 + \nu\cdot\chi\left(\omega\right)\right|^2\right),$ where $\kappa = \nu +1$ is the expected number of shocks per primary shock.
\end{enumerate}
(both in the sense of LHS converging in distribution to RHS as $M\rightarrow \infty$, with $\lambda$ fixed, conditioning on $M$, and $M$ increasing.)
\end{Prop}
\begin{proof}
(i) follows from the discussion in Appendix~\ref{sec:app-schuster-test}.\\
(ii) follows from the discussion in Appendix~\ref{sec:app-schuster-test-aft}.
\end{proof}

As per Proposition~\ref{Prop:aftershockschuster}~(ii), we note that the scaling constant of the Schuster spectrum, equal to the parameter of the $\Exp$ distribution versus $\omega$, is no longer known a-priori, due to $\chi\left(\omega\right)$ and $\kappa$ being unknown. In addition, it has a dependence on $\omega$.\\
This means that, in the presence of an aftershock process, it is no longer appropriate to normalise the Schuster periodogram by the number of earthquakes $N$.\\

If it is possible to say with certainty which earthquakes are main shocks and which are aftershocks, $\chi$ and $\kappa$ may be directly estimated, as outlined in Section~\ref{sec:estimation}. Otherwise is therefore not practical in most cases to use equation~\eqref{eq:prop1} directly, especially as the labelling of primary vs aftershocks may not even  be identifiable.

A semi-heuristic correction based on the above theoretical insights may nevertheless be derived, subject to the following (semi-heuristic) assumptions:
\begin{enumerate}
\itemsep-0.2em
\item[(a)] the observation window is large, in comparison to the aftershock process (mathematically: $\Var (Z) \ll 1$ for $Z$ as in Lemma~\ref{Lem:aftvar})
\item[(b)] $\chi$ varies smoothly in $\omega$, and deviations from the null appear only in isolated peaks.
\item[(c)] $D(\omega)$, for different close-by values of $\omega$, is close to being independent
\end{enumerate}

Subject to the above, we can estimate the rate parameter of the Schuster spectrum by regression on the sample estimate of $\frac{1}{N}D^2(\omega)$ against $\omega$, i.e., subject to smoothness regularization on $\omega$. Periodicity will then appear as an outlier in this $\omega$.\\
We opt for using an off-shelf class of regression methods with the above behaviour, robust quantile regression, regressing at the $(1-\exp(-1))$-quantile which is always identical with the mean of an exponential distribution.

The full algorithm for our modified Schuseter spectrum test is as follows:

\begin{algorithm2e}[H]
	\SetAlgoLined
	\SetKwInOut{Input}{Input}\SetKwInOut{Output}{Output}
	\Input{occurrence times, $t_1,\dots,t_N$ \\
     periods to consider, $k_1,\dots, k_K$}
	\Output{Schuster p-values $p_1,\dots, p_K$ for these periods }
 compute $D^2_j \gets \left|\sum_{i=1}^{N}{\exp{\left(\frac{2 \pi t_i}{k_j}\right)}}\right|^2$, for $j$ in $\{1,\dots, K\}$\;
 initialize quantile regression model \texttt{QRM}, with quantile $1-\exp(-1)\approx 0.63$\; \label{Alg1.QRM.init}
 fit \texttt{QRM} to sample of covariates $k_1,\dots, k_K,$ targets $D^2_1,\dots, D^2_K$ \;
 use \texttt{QRM} to predict quantiles $\widehat{D}^2_1,\dots, \widehat{D}^2_K$ at $k_1,\dots, k_K$\;\label{Alg1.QRM.predict}
 compute $p_j \gets \exp\left(- D^2_j/ \widehat{D}^2_j\right)$, for $j$ in $\{1,\dots, K\}$\;\label{Alg1.norm}
 return $p_1,\dots, p_K$
 \caption{Modified Schuster Spectrum Test}
\end{algorithm2e}
\vspace{0.5cm}
The algorithm, as described, returns unadjusted p-values, $p_i$ being a p-value for periodicity at period length $k_i$, subject to known asymptotics and heuristic assumptions made. This is computed for an earthquake catalogue with observation times $t_1,\dots, t_N$.

In set-up, our algorithm follows closely the SST procedure described in \cite{Ader2013} - both compute the p-value at a frequency $\omega$ based on the exponential asymptotic of the Schuster $D^2$, normalizing by its expectation/rate parameter $D^2(\omega)$. The main difference arises through the frequency dependency of $D^2(\omega),$ see Proposition~\ref{Prop:aftershockschuster}. We address this by estimating the normalization factor $\widehat{D}^2_j$, then use this in line~\ref{Alg1.norm} in place of the factor $N$ in the original SST algorithm.\\

As explained in the discussion above, we opt to conduct this estimation by a quantile regression algorithm, on the calculated Schuster spectrum, in lines~\ref{Alg1.QRM.init} to~\ref{Alg1.QRM.predict}. This strategy seems to work well with off-shelf quantile regressors, if the periods for the spectrum are equally spaced in frequency (not length) domain.

In our own implementation, for the quantile regression method \texttt{QRM} in the algorithm, we use the B-spline quantile regression algorithm from the R \url{cobs} package~\cite{COBS}, version 1.3-4, which implements the quantile regression method described in~\cite{Ng2007}. We interface the \texttt{cobs} quantile regressor subject to the constraint that for very short periods the expected value of $D^2_j$ approaches $N$, as derived in Section \ref{sec:distSpec}. Otherwise, we run \texttt{cobs} with unchanged default parameter settings. In principle, further constrains can be placed on the spline fit, for example if partial information is know about the underlying physical process, however we have chosen not to apply any such constraints to keep this method as general as possible.

In all our experiments below, the period length $k_1,\dots, k_K$ are taken to be equally spaced in the frequency domain, with $k_1 = 1$ day and $k_K = 5$ years, and $K=\left\lceil\frac{k_K^{-1}-k_1^{-1}}{T_N-T_1}\right\rceil,$ following~\cite{Ader2013}. That is, $k_i^{-1} = k_1^{-1} + i\cdot \frac{k_K^{-1}-k_1^{-1}}{K-1}$ for $i=1\dots K$.

Significances in the reported results will be subject to  multiple testing correction if multiple periods lengths are considered. For simplicity, we apply Bonferroni correction which is a procedure developed by~\cite{Dunn1961}. More precisely, if we apply our test to $K$ different periods, then the corrected p-value for period length $k_j$ is $K\cdot p_k$. We choose this method due to its simplicity in interpreting the graphically presented results and scrutinizing the impact of the correction; of course other, more efficient methods to control family-wise error rate (such as Holm/Bonferroni or Hochberg) could be applied here.

\subsection{The MSST applied to an example process}

\begin{figure}[H]
\includegraphics[width = \textwidth]{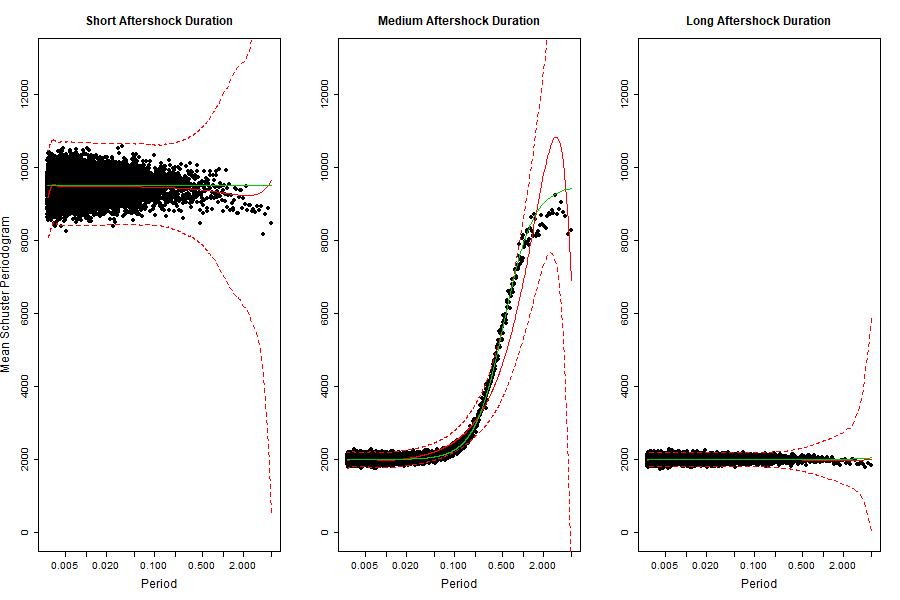}
\caption{This figure shows the results of calculating the Schuster periodogram, $D^{2}_k$ for 1000 simulated earthquake catalogues. The black points show the average of the calculated values of $D^{2}_k$ over the simulations. The green line shows the theoretical mean of $D^{2}_k$ as calculated from equation \eqref{eq:prop1}. The solid red line shows the mean of the fitted values obtained by fitting a spline to each of the 1000 simulations, the dashed red lines show the pointwise 95\% quantiles of the fitted values.}
\label{fig:sims}
\end{figure}

A demonstration of the MSST on simulated data can be seen in Figure \ref{fig:sims}. In this plot we have simulated earthquake catalogues which include aftershocks. The aftershock process follows that defined in equation \eqref{eq:afm} with $\tau$ equal to either $1\times 10^{-5}$ years, 1/12 years or 10 years. This represents the extremely short and long aftershock durations discussed in Section \ref{sec:newTest} as well as a medium duration which is more typical of what is seen in real data. Here we have set the parameters $\nu_0=500$ and $\nu = 3$, this corresponds to an expected value of $N$ of 2000. The periods we consider range from a minimum of 1 day up to 5 years.

Figure \ref{fig:sims} shows the results of this simulation. We can see from these plots that in general the true value of the Schuster periodogram lies within the 95\% uncertainty rage of our estimate. We do note that for very long periods the uncertainty range becomes very large, this is due to the spline fit being poorly constrained in this region due to a lack of observable periods. This could be reduced by either adding some constrains, as discussed above, or where possible increasing the observation period. We also note that for long periods the mean of the simulated periodograms deviates form the theoretical expected value given by equation \eqref{eq:prop1}. This is likely due to a finite sample effect caused by the finite observation window. While this is noted we are as yet unable to properly quantify this effect.

\newpage
\section{Results}\label{sec:results}

In this Section we present some results of applying the MSST to both simulated data in Section \ref{sec:simData} and to real data in Section \ref{sec:realData}. In both cases we present the results of our method alongside those of the current SST. In the case of simulated data this allows us to understand the impact of aftershocks, and to investigate how each method is able to perform in a situation where we know exactly the underlying earthquake generating model and therefore we know what the ``optimal'' results are. In the case of real data we demonstrate how the two models perform outside of the idealised setting of simulated data.

\subsection{Simulated Data - Simple Aftershock Process}
\label{sec:simData}

In this section we present the results of applying the MSST to simulated data. We describe simulation procedures for four different data generative scenarios (A,B,C,D), which are later used to generate independent replicates of experiments, in which one particular scenario is independently re-sampled. One replicate, in each of the four scenarios, is generated in the following way:

\begin{enumerate}[label = {\bf (\Alph*)}]
\item No Aftershocks and No Seasonality:
	\begin{enumerate}[label = \arabic*.]
		\item We simulate a catalogue with $N$ earthquakes where $N$ is taken from a Poisson distribution with mean $\nu_{0} = 2000$.
		\item The occurrence times of the earthquakes are sampled i.i.d.~and uniformly from the interval between $0$ and 50 years.
	\end{enumerate}
\item Aftershocks and No Seasonality:
	\begin{enumerate}[label = \arabic*.]
		\item The number $M$ of primary shocks is sampled from a Poisson distribution with mean $\nu_{0} = 500$.
		\item The time of the each primary shock is sampled uniformly (i.i.d.) between $0$ and 50 years.
		\item For each primary shock, we additionally simulate $A_m$ aftershocks, where $A_m$ are sampled i.i.d.~from a Poisson distribution with mean $\nu = 3$.
		\item The waiting times for each aftershock, after the main shock, are sampled i.i.d.~from an exponential distribution with half-life of $\tau = 1$ month.
	\end{enumerate}
\item No Aftershocks and Seasonality:
	\begin{enumerate}[label = \arabic*.]
		\item The total number of shocks $N$ is sampled from a Poisson distribution with mean $\nu_{0} = 2000$.
		\item Occurrence times of the shocks are sampled i.i.d.~from the distribution with probability density function proportional to $1+\alpha \mbox{ sin}\left(2 \pi t/t_s\right)$, as a function in $t$, for $t\in [0, 50]$ years, where $\alpha = 0.5$ and $t_s = 1$ year.
	\end{enumerate}
\item Aftershocks and Seasonality:
	\begin{enumerate}[label = \arabic*.]
		\item First the number $M$ of primary shock is sampled from a Poisson distribution with mean  $\nu_{0} = 500$.
		\item Occurrence times of the primary shocks are sampled i.i.d.~from the distribution with probability density function proportional to $1+\alpha \mbox{ sin}\left(2 \pi t/t_s\right)$, as a function in $t$, for $t\in [0, 50]$ years, where $\alpha = 0.5$ and $t_s = 1$ year.
		\item For each primary shock, we additionally simulate $A_m$ aftershocks, where $A_m$ are sampled i.i.d.~from a Poisson distribution with mean $\nu = 3$.
		\item The waiting times for each aftershock, after the main shock, are sampled i.i.d.~from an exponential with half-life of $\tau = 1$ month.
	\end{enumerate}
\end{enumerate}

\begin{figure}[H]
\centering
\includegraphics[width = 0.9\textwidth]{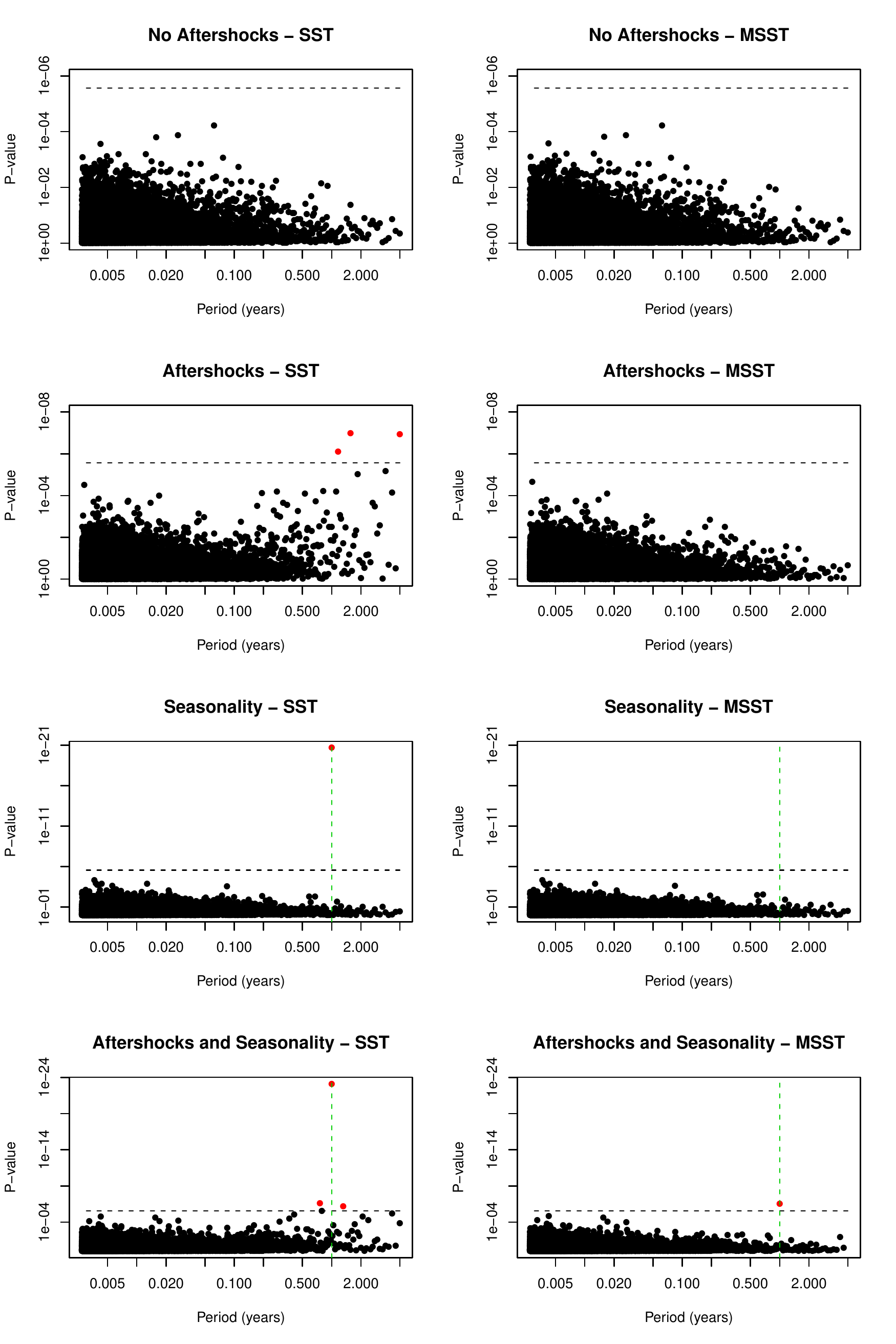}
\caption{\small{Each plot shows the results of applying either the standard SST, left hand column, or our MSST, right hand column, to single simulated catalogues from four different scenarios. In each case the y axis shows the p-value for each period considered. The black line shows the Bonferroni corrected 5\% significance threshold for $K = 18241$ periods. Each point on the plots show the calculated Schuster spectrum for each period. Points in red are above the significance threshold while those in black are below. The green vertical line shows the period of seasonality.}}
\label{fig:examples}
\end{figure}

\begin{figure}[H]
\centering
\includegraphics[width = 0.9\textwidth]{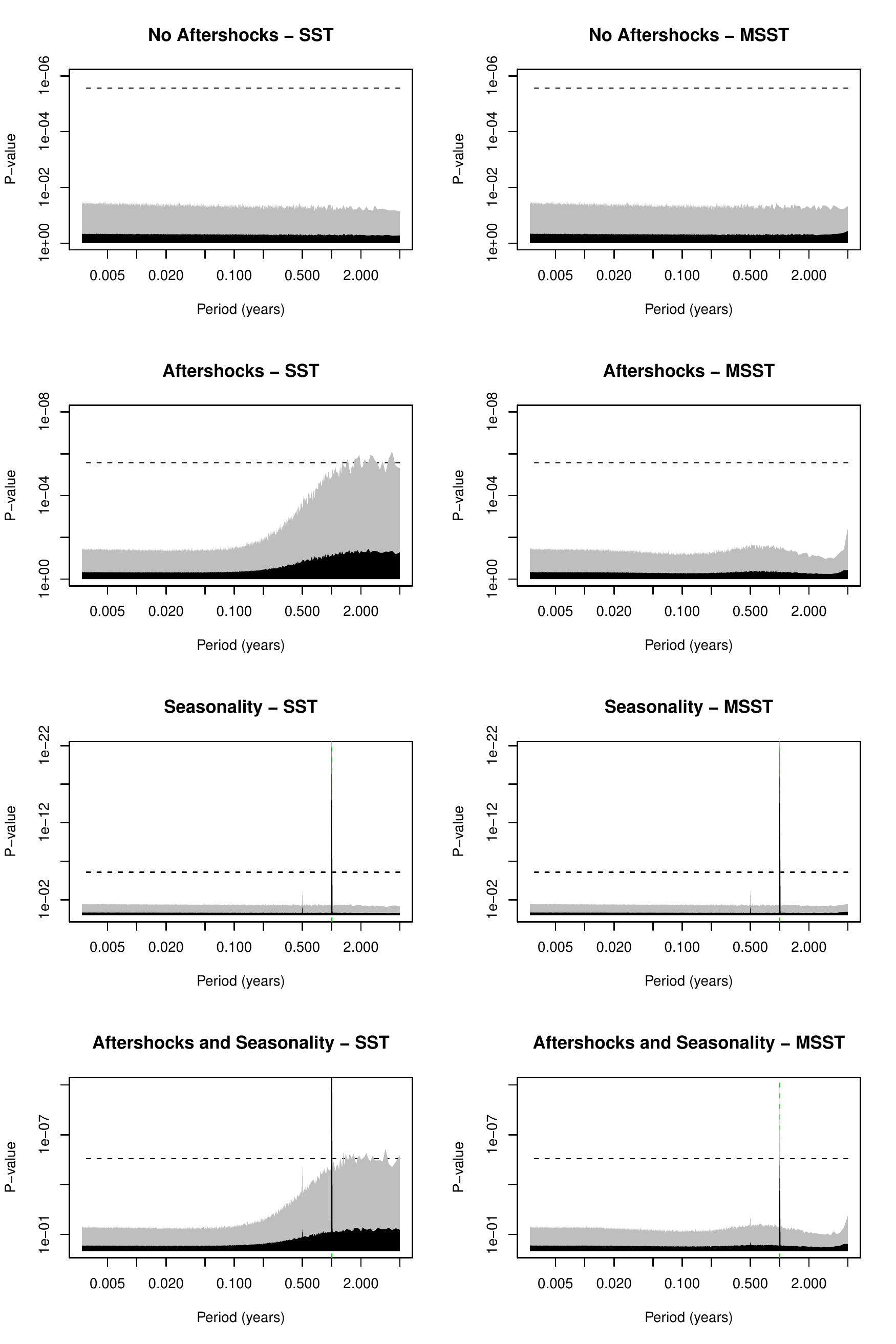}
\caption{\small{Each plot shows the results of applying either the standard SST, left hand column, or our MSST, right hand column, to four different scenarios for 1000 simulations. In each case the y axis shows the p-value for each period considered. The black line shows the Bonferroni corrected 5\% significance threshold for $K = 18241$ periods. For each case 95\% of the calculated spectrum values from those experiments were within the gray shaded area and 50\% are within the black area. The green vertical line shows the period of seasonality.}}
\label{fig:examplesShading}
\end{figure}

The results of applying both the existing SST and our modified version to the four different scenarios are shown in Figures \ref{fig:examples} and \ref{fig:examplesShading}. Here again we consider periods between 1 day and 5 years with steps defined at the end of Section~\ref{sec:adaptedTest} for both tests. Looking at these figure we see that for the two cases where there are no aftershocks both methods perform to a similar level. Neither shows any positive results for the single catalogue run, Figure \ref{fig:examples}, for the case of no aftershocks and no seasonality. Both show a positive result for the period corresponding to the seasonality in the case of no aftershocks with seasonality. As intended there are larger differences when the catalogue contains aftershocks. For these cases the existing SST shows a number of positive results for periods which do not correspond to a seasonal pattern. The gray shaded area in Figure \ref{fig:examplesShading}, equivalent to the 95\% quantile for the calculated spectrum, reinforce this with a significant increase for periods greater than the aftershock parameter $\tau = 1$ month. This increase is greatly reduced by our method.

We can also extend this simulation to quantify the false discovery rates for the two methods. We do this by repeatedly simulating the earthquake catalogues and applying both methods, we can then count the proportion of cases where each method returns a positive test result. We do this for 1000 simulations under scenarios A and B defined above. In both of these cases there is no true seasonal component and so any positive results should be counted as a false positives. We conduct both tests at a 5\% significance level and so would expect a false discovery rate of 5\%. The results of this are shown in Table \ref{tab:FP}. Looking at this table we see that for the case of no aftershocks both methods have a false discovery rate of just above 7.4\%. For the case where there are aftershocks our method gives a very similar false discovery rate, however the current method shows a much higher false discovery rate of close to 97\%.

\begin{table}[H]
	\caption{False positive rates for the current Schuster spectrum test and our new method. This is calculated for 1000 simulations both with and without aftershocks.}\label{tab:FP}
	\centering
	\begin{tabular}{|l|ll|}
		\hline
		Aftershocks & SST & MSST \\
		\hline
		No          &      0.063      &    0.074       \\
		Yes         &      0.971      &    0.070       \\
		\hline
		\end{tabular}
\end{table}

\subsection{Simulated Data - Recursive Aftershock Process}
\label{sec:simDataRecursive}

The aftershock process used within Section \ref{sec:simData} is a simple, one generational, process which is not truly representative of the more complex processes which can be observed in nature. We use this simple aftershock process as it allowed us to derive an analytical expression for the Schuster distance in Section \ref{sec:distSpec}. The modified test described in Section \ref{sec:adaptedTest} however does not make strong assumptions on the particular aftershock process and so will be robust to different aftershock processes. In this section we demonstrate this by applying both the SST and our MSST to simulated data which includes both multi generational aftershocks, i.e. aftershocks can themselves produce further aftershocks, and also a magnitude depended rate of aftershock generation. These catalogues are simulated in the following way,

\begin{enumerate}
	\item The number of primary shocks, $M$, is sampled from a Poisson distribution with mean $\nu_0 = 500$.
	\item Occurence times of the primary shocks are sampled in one of two ways,
	\begin{enumerate}
		\item  In the case of no seasonlity: Uniformly (i.i.d.) between 0 and 50 years in the case of no seasonality.
		\item In the case of seasonality: i.i.d.~from the distribution with probability density function proportional to $1+\alpha \mbox{ sin}\left(2 \pi t/t_s\right)$, as a function in $t$, for $t\in [0, 50]$ years, where $\alpha = 0.5$ and $t_s = 1$ year.
	\end{enumerate}	
	\item For each main shock we simulate a magnitude, $\mu_{i}$ from an exponential distribution with rate $b=1$.
	\item The number of aftershocks for each main shock follows a Poisson distribution with mean of $\nu\exp\left[{a (\mu_i - \mu_{\mbox{min}})}\right]$ where $\nu = 0.75$ is the aftershock rate, $\mu_{\mbox{min}} = 2$ is the magnitude of completeness and $a = 1$ is the parameter controlling the degree of magnitude dependence.
	\item Magnitudes of these aftershocks are then simulated from an exponential distribution with mean $b = 1$ while the waiting times for each aftershock, after the main shock, are sampled i.i.d.~from an exponential with half-life of $\tau = 1$ month.
	\item These aftershocks are then treated as the main shocks and are able to produce further aftershocks of their own following steps 4 and 5.
	\item This process continues until no further aftershocks are produced.
\end{enumerate}

\begin{figure}[H]
\centering
\includegraphics[width = 0.9\textwidth]{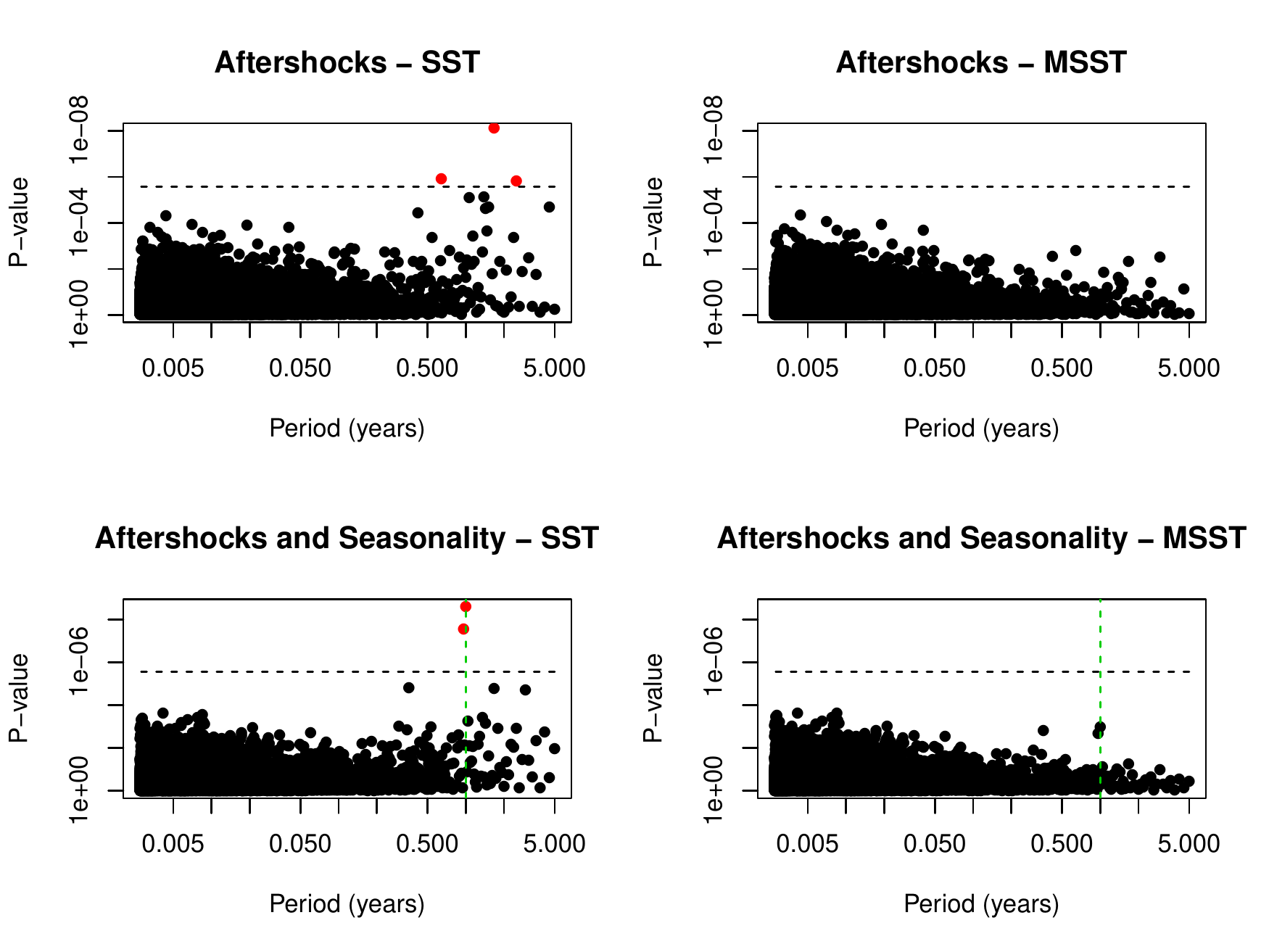}
\caption{\small{Each plot shows the results of applying either the standard SST, left hand column, or our MSST, right hand column, to single simulated catalogues from two different scenarios with the more complex aftershock process. In each case the y axis shows the p-value for each period considered. The black line shows the Bonferroni corrected 5\% significance threshold for $K = 18241$ periods. Each point on the plots show the calculated Schuster spectrum for each period. Points in red are above the significance threshold while those in black are below. The green vertical line shows the period of seasonality.}}
\label{fig:examplesRecursive}
\end{figure}

\begin{figure}[H]
\centering
\includegraphics[width = 0.9\textwidth]{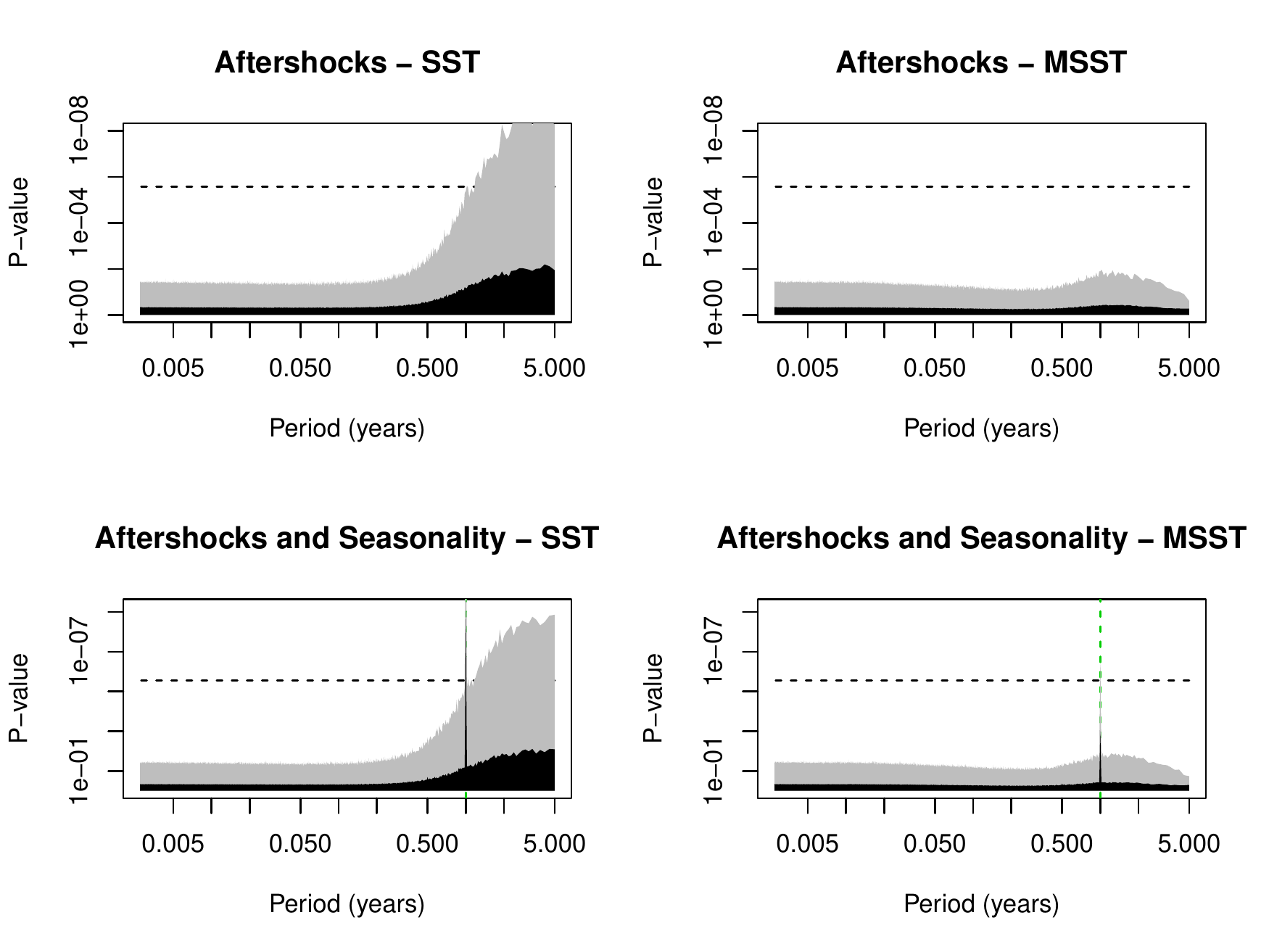}
\caption{\small{Each plot shows the results of applying either the standard SST, left hand column, or our MSST, right hand column, to two different scenarios with the more complex aftershock process for 1000 simulations. In each case the y axis shows the p-value for each period considered. The black line shows the Bonferroni corrected 5\% significance threshold for $K = 18241$ periods. For each case 95\% of the calculated spectrum values from those experiments were within the gray shaded area and 50\% are within the black area. The green vertical line shows the period of seasonality.}}
\label{fig:examplesShadingRecursive}
\end{figure}

In Figures \ref{fig:examplesRecursive} and \ref{fig:examplesShadingRecursive} we can see the results of rerunning scenarios B and D from Section \ref{sec:simData} with the aftershock process described above. Looking at Figure \ref{fig:examplesRecursive} we see that  the SST gives false positive results in the case of Aftershocks but non seasonality whereas the MSST does not. In the scenario where there is both seasonality and aftershocks the SST gives a positive result for the period of seasonality however it also gives an additional positive result for a neighbouring period. The MSST does not show positive result for any frequency. Looking at Figure \ref{fig:examplesShadingRecursive}, and comparing with Figure \ref{fig:examplesShading}, we can see that this aftershock process has had a larger effect on the SST with a much larger chance of false positive results in both scenarios. The MSST however is still able to control the false positive rate. For this case we do see that when seasonality is present the power of the test to discover the true seasonality is reduced, likely this is as a result of the correcting for the more pronounced aftershock process.  

\subsection{Seasonal seismicity within the New Madrid Seismic Zone}
\label{sec:realData}
\begin{figure}
\centering
\includegraphics[width = 0.9\textwidth]{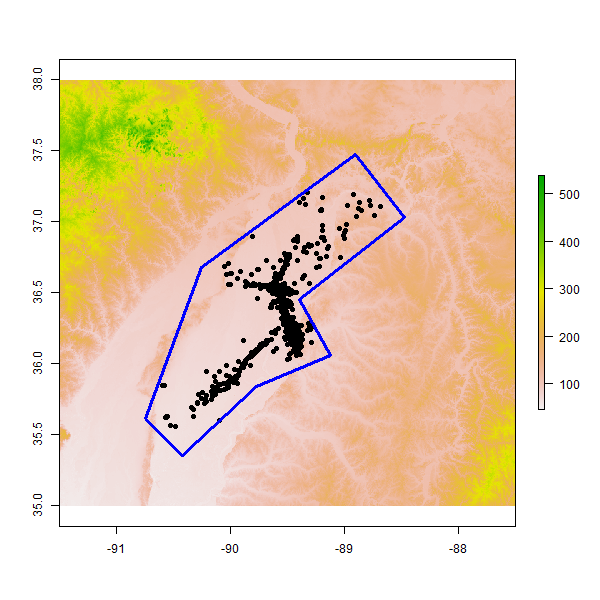}
\caption{\small{Earthquake epicentres (black dots) within the New Madrid Seismic Zone (blue polygon) from the CERI catalogue \citep{CERI2019} for the period 1/1/2000 to 1/1/2016 and magnitudes $M \ge 1.4$. Colours denote surface elevation in meters and map coordinates are in degrees of longitude and latitude.}}
\label{fig:newMadridMap}
\end{figure}

In this section we apply our method to an observed earthquake catalogue from the New Madrid Seismic Zone for the period 1/1/2000 to 1/1/2016 (Figure~\ref{fig:newMadridMap}). This is an intra-plate region with very small secular stressing rates. Annual and multi-annual stress variations are driven by seasonal changes in the terrestrial water mass.
This catalogue was analysed by \cite{Craig2017} who found clear evidence for yearly seasonal patterns in the earthquake occurrence rate. They did this by considering the difference between observed rates in two 4 month periods, January to April and July to October, and comparing this to simulated catalogues with the same magnitude-frequency distribution. This approach is more specific than the SST as only considers seasonality at yearly frequency and assumes the phase of seasonality is aligned with the chosen months. It does however make a similar assumption of independence between events and so requires prior declustering. For this data set we have taken the magnitude of completeness, the minimum magnitude above which we expect the catalogue to be complete, as $M_c=1.9$. This differs from \cite{Craig2017} who selected $M_c=1.4$ based on the apparent onset of under-sampling in an magnitude versus exceedance count plot. However, for $M_c=1.4$ we found significant evidence for daily periodicity in the catalogue consistent with a systematic day-night variation in event detectability indicating $M_c > 1.4$. Repeating this daily periodicity analysis with progressively larger $M_c$ values we find the day-night effect first disappears for $M_c=1.9$.

\paragraph{Declustering: } Since the SST is sensitive to aftershocks one approach which can be taken is to first decluster the catalogue and then apply the seasonality test. In theory this should remove the aftershocks from the catalogue whilst preserving the seasonal pattern in the main shocks. In practice it can be very difficult to apply declustering without making strong assumptions about the underlying earthquake generating process. For this example we apply the following simple declustering to the New Madrid catalogue, using a standard method based on the space–time linked windows \citep[e.g.][]{Reasenberg1985}.

\begin{itemize}
\item Iterate through the catalogue in order of occurrence time.
\item For the current earthquake any subsequent earthquake is considered an aftershock if:
	\begin{enumerate}
	 	\item It occurs within 5 days of the time of the current earthquake.
	 	\item It occurs within 10km of the current earthquake.
	\end{enumerate}
\item Remove all earthquakes fitting the above criteria and continue iterating through the catalogue, ignoring those which have already been identified as aftershocks.
\end{itemize}
Stochastic declustering methods based on the ETAS model for spatial-temporal clustering \citep{Zhuang2002} yield a sample of many alternatively declustered catalogues suitable for probabilistic seismic hazard analysis but not suited to the SST that requires a single catalogue.

We now apply three tests to the New Madrid data,
\begin{enumerate}
	\item The MST applied to the full catalogue
	\item The original SST applied to the full catalogue.
	\item The original SST applied to the declustered catalogue.
\end{enumerate}

\begin{figure}
\begin{subfigure}{0.9\textwidth}
\centering
\includegraphics[width = 0.9\textwidth]{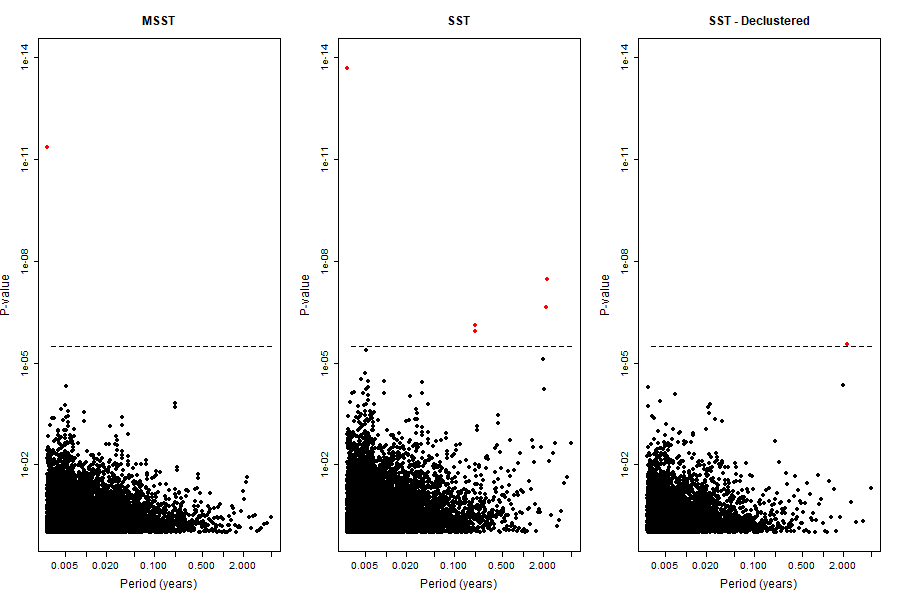}
\caption{\small{Minimum magnitude $M_c = 1.4$. }}
\label{fig:newMadridExampleM14}
\end{subfigure}

\begin{subfigure}{0.9\textwidth}
\centering
\includegraphics[width = 0.9\textwidth]{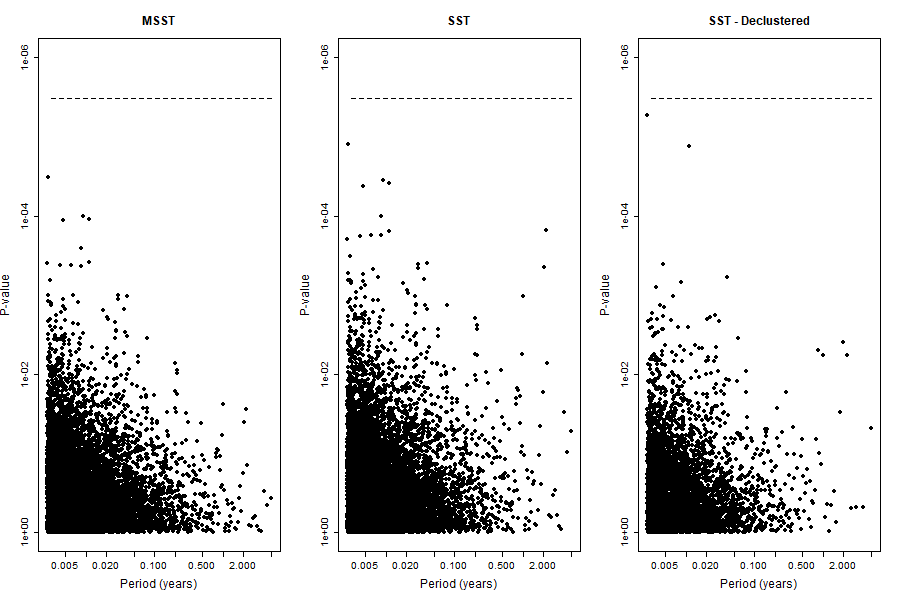}
\caption{\small{Minimum magnitude $M_c = 1.9$.}}
\label{fig:newMadridExampleM19}
\end{subfigure}
\caption{\small{This figure shows the results of applying both the original SST, left, and our MSST, right to the New Madrid earthquake catalogue for two choices of minimum magnitude, $M_c$. The periods tested are plotted on the x-axis and the p-values of each test are plotted on the y-axis. In both cases the dashed line shows the Bonferroni corrected 5\% significance threshold for $K = 15243$ periods.}}
\label{fig:newMadridExample}
\end{figure}

As with the simulated example we compute the spectrum at equally spaced frequencies between inverse 1 day and 5 years, as described at the end of Section~\ref{sec:adaptedTest}.
The results of this can be seen in Figure \ref{fig:newMadridExample}. Looking at Figure \ref{fig:newMadridExampleM14}, showing results for magnitudes greater than or equal to 1.4, we see that when we apply both tests to the data without declustering we find significant seasonality with a period of 1 day. This we attribute to a sampling bias in the catalogue which is an indication that the catalogue is not complete and the minimum magnitude if not large enough. It is worth noting that the test applied to declustered data does not show significant daily periodicity, this could indicate that the declustering process has destroyed this daily seasonal signal.

Looking at Figure \ref{fig:newMadridExampleM19}, showing results for magnitudes greater than or equal to 1.9, the first thing we note is that in all cases we do not find any significant seasonality. While this may appear to contradict the findings of \cite{Craig2017} we note that this may be explained by the differences in testing procedures used. Comparing the results of the individual tests in Figure \ref{fig:newMadridExampleM19} we note that while there are no significant results for the original Schuster Spectrum the p-values for longer periods are noticeably lower than for our adapted version. This is likely a result of aftershock effects as they do not coincide with any expected seasonal pattern. Declustering the data has reduced this effect to some extent however not to the extent of our test which does not require a separate declustering algorithm. Reducing the magnitude of completeness to $M_c=1.4$, as selected by \cite{Craig2017} does not change the results for the MSST, we still find no statistical evidence for seasonal seismicity, although we do find significant daily seismicity variations.

\subsection{Seasonal seismicity within the Himalayas}
\label{sec:himalayaExample}

\begin{figure}
\centering
\includegraphics[width = 0.9\textwidth]{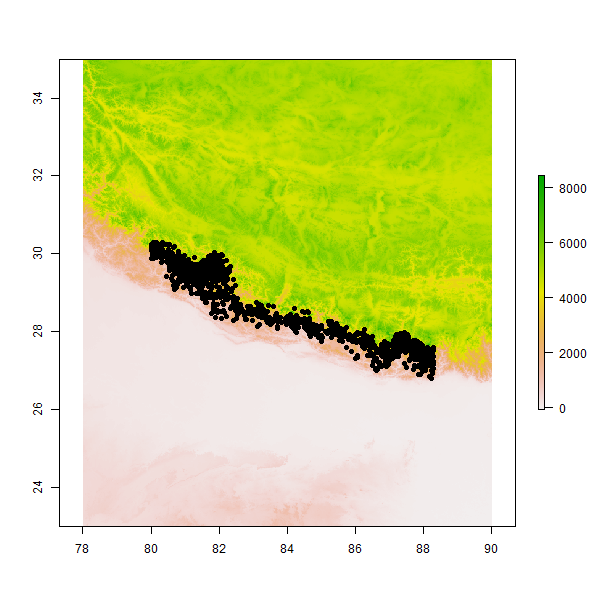}
\caption{\small{Earthquake epicentres (black dots) within the Nepal midcrustal cluster for the period 1/1/1995 to 1/1/2001 and magnitudes $M \ge 3$. Colours denote surface elevation in meters and map coordinates are in degrees of longitude and latitude.}}
\label{fig:himalayaMap}
\end{figure}

In this section we present the results of applying the SST and our MSST to a data set from the Himalayas. This data was also analysed by \cite{Ader2013} who also applied the SST and found evidence of seasonal patterns at several different periods. The data used is restricted to events in the midcrustal cluster of seismicity with a minimum magnitude of 3 and for the years 1995-2001. This again follows the choices of \cite{Ader2013}. A map of these events is shown in \ref{fig:himalayaMap}.

 In their paper \cite{Ader2013} first apply declustering to this data using the approach described in \cite{Bollinger2007}. For our comparison we first analyse this data using the SST and MSST on the data without declustering. We then apply the SST and MSST to two sets of data which have been declusterd following the method of\cite{Bollinger2007}. First with the distance parameter $D\leq20$km and secondly with $D\leq80$km. This allows us to show how consistent the results are in relation to the choice of parameter in the declustering step. The declustered data was kindly provided to us by the authors of \cite{Adder2013} which should allow for a more direct comparison with their results.

\begin{figure}
\centering
\includegraphics[width = 0.9\textwidth]{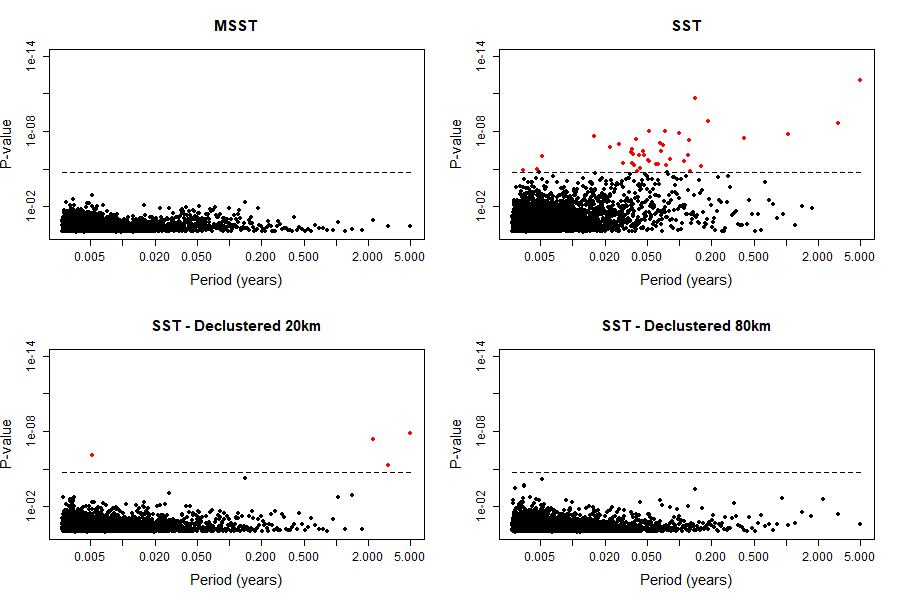}
\caption{\small{This figure shows the results of applying both our modified MSST, top left, and the original SST, top right right, to the Nepal earthquake catalogue. Additionally the SST is also applied to the declustered catalogue for two choices of parameter $D\leq 20$km and $D\leq80$km. The periods tested are plotted on the x-axis and the p-values of each test are plotted on the y-axis. In both cases the dashed line shows the Bonferroni corrected 5\% significance threshold for $K = 2920$ periods.}}
\label{fig:himalayaExample}
\end{figure}

The results can be seen in Figure \ref{fig:himalayaExample}. We can see that the MSST does not find any significant seasonal patter in the data. When applied to data without declustering the SST finds lots of significant seasonal periods, however this is very likely due to false positives driven by the aftershock process. For the declustered data the SST shows some significant periods for the choice of parameter $D\leq20$km but none for $D\leq80$km. This is consistent with the findings of \citep{Ader2013} who only consider the first case. These results demonstrate the the declustering step can have a significant impact on the results of the analysis when using the SST. The MSST however does not require a separate declustering step and so is not sensitive to this choice.

\section{Conclusion}\label{sec:conclusions}

In this paper we have shown that, in the presence of aftershocks, the expected value of the Schuster periodogram is given by equation \eqref{eq:prop1}. This result demonstrates why the Schuster spectrum test introduced by \cite{Ader2013} gives an elevated false positive rate when applied to earthquake catalogues which contain aftershocks. Furthermore in Section \ref{sec:newTest} we are able to give some physical meaning to the terms in equation \eqref{eq:afm} in terms of the expected number of main shocks and aftershocks as well as the characteristic aftershock function. We also describe certain edge cases where this formula can be simplified.

In principle these theoretical result could be used to construct an improved hypothesis test which accounts for the effect of aftershocks. However in general it is not possible to estimate the unknown parameters in equation \eqref{eq:prop1} without making strong assumptions about the underlying physical process. In Section \ref{sec:adaptedTest} we described an alternative procedure which is to first make a non-parametric estimate of the expected value of the Schuster spectrum, $D^2(k)$ using the calculated values of the Schuster spectrum, $D^2_k$. In Section \ref{sec:results} we demonstrated that our improved test is better able to control the false positive rate for catalogues containing aftershocks, while still detecting true periodicity.

In conclusion we believe that our Modified Schuster Spectrum test has been shown to outperform the standard test for earthquake catalogues which contain aftershocks. This comes with the additional advantage that we do not require a separate declustering step to remove aftershocks. We believe this has a wide ranging applicability because we expect most observed earthquake catalogues to contain aftershocks.

\section*{Authors' contributions}
SJB stated the original, high-level research question. The technical content was developed mostly by FJK and TP in discussions. FK carried out the methodological literature review, SJB carried out the domain literature review. FK proved most of the main theoretical results, e.g., Propositions~\ref{prop:exd} and~\ref{Prop:aftershockschuster}. TP devised the algorithm for the aftershock adapted test. SJB planned the experiments on real world data. TP implemented the algorithm and carried out the experiments. The manuscript was written by all authors, split roughly across contribution areas; the entire manuscript was reviewed by all authors, and all authors are responsible for its content.

\section*{Acknowledgements}
We gratefully acknowledge our colleagues from Nederlandse Aardolie Maatschappij Jan van Elk, Dirk Doornhof and from Shell Global Solutions International, Keimpe Jan Nevenzeel, Stijn Bierman, Phil Jonathan for their ongoing support during this study. We thank Prof. Jean-Philippe Avouac for helpful discussions and suggestions regarding the development and application of the Schuster and the Schuster Spectrum tests.

\clearpage
%\bibliography{ref}
\bibliography{ref,SeasonalityPaper}

\begin{thebibliography}{}

\bibitem[Ader and Avouac, 2013]{Ader2013}
Ader, T.~J. and Avouac, J.-P. (2013).
\newblock {Detecting periodicities and declustering in earthquake catalogs
  using the Schuster spectrum, application to Himalayan seismicity}.
\newblock {\em Earth and Planetary Science Letters}, 377-378:97--105.

\bibitem[Ader et~al., 2014]{Ader2014}
Ader, T.~J., Lapusta, N., Avouac, J.-P., and Ampuero, J.-P. (2014).
\newblock {Response of rate-and-state seismogenic faults to harmonic
  shear-stress perturbations}.
\newblock {\em Geophysical Journal International Geophys. J. Int},
  198:385--413.

\bibitem[Beeler and Lockner, 2003]{Beeler2003a}
Beeler, N.~M. and Lockner, D.~A. (2003).
\newblock {Why earthquakes correlate weakly with the solid Earth tides: Effects
  of periodic stress on the rate and probability of earthquake occurrence}.
\newblock {\em Journal of Geophysical Research}, 108(B8).

\bibitem[Bettinelli et~al., 2008]{Bettinelli2008}
Bettinelli, P., Avouac, J.~P., Flouzat, M., Bollinger, L., Ramillien, G.,
  Rajaure, S., and Sapkota, S. (2008).
\newblock {Seasonal variations of seismicity and geodetic strain in the
  Himalaya induced by surface hydrology}.
\newblock {\em Earth and Planetary Science Letters}, 266(3-4):332--344.

\bibitem[Bollinger et~al., 2007]{Bollinger2007}
Bollinger, L., Perrier, F., Avouac, J.~P., Sapkota, S., Gautam, U., and Tiwari,
  D.~R. (2007).
\newblock {Seasonal modulation of seismicity in the Himalaya of Nepal}.
\newblock {\em Geophysical Research Letters}, 34(8).

\bibitem[Bourne and Oates, 2017]{Bourne2017a}
Bourne, S.~J. and Oates, S.~J. (2017).
\newblock {Extreme threshold failures within a heterogeneous elastic thin-sheet
  and the spatial-temporal development of induced seismicity within the
  Groningen gas field}.
\newblock {\em Journal of Geophysical Research: Solid Earth},
  122:10,299--10,320.

\bibitem[Bourne et~al., 2018]{Bourne2018}
Bourne, S.~J., Oates, S.~J., and Elk, J.~V. (2018).
\newblock {The exponential rise of induced seismicity with increasing stress
  levels in the Groningen gas field and its implications for controlling
  seismic risk}.
\newblock {\em Geophysical Journal International}, 213:1693--1700.

\bibitem[Bourne et~al., 2014]{Bourne2014a}
Bourne, S.~J., Oates, S.~J., van Elk, J., and Doornhof, D. (2014).
\newblock {A seismological model for earthquakes induced by fluid extraction
  from a subsurface reservoir}.
\newblock {\em Journal of Geophysical Research: Solid Earth},
  119(12):8991--9015.

\bibitem[Brockwell and Davis, 2013]{Brockwell2013}
Brockwell, P. and Davis, R. (2013).
\newblock {\em Time Series: Theory and Methods}.
\newblock Springer Series in Statistics. Springer New York.

\bibitem[CERI, 2019]{CERI2019}
CERI (2019).
\newblock {New Madrid Earthquake Catalog}.

\bibitem[Chanard et~al., 2019]{Chanard2019}
Chanard, K., Nicolas, A., Hatano, T., Petrelis, F., Latour, S., Vinciguerra,
  S., and Schubnel, A. (2019).
\newblock {Sensitivity of Acoustic Emission Triggering to Small Pore Pressure
  Cycling Perturbations During Brittle Creep}.
\newblock {\em Geophysical Research Letters}, 46(13):7414--7423.

\bibitem[Christiansen et~al., 2005]{Christiansen2005}
Christiansen, L.~B., Hurwitz, S., Saar, M.~O., Ingebritsen, S.~E., and Hsieh,
  P.~A. (2005).
\newblock {Seasonal seismicity at western United States volcanic centers}.
\newblock {\em Earth and Planetary Science Letters}, 240(2):307--321.

\bibitem[Cochran et~al., 2004a]{cochran2004earth}
Cochran, E.~S., Vidale, J.~E., and Tanaka, S. (2004a).
\newblock Earth tides can trigger shallow thrust fault earthquakes.
\newblock {\em Science}, 306(5699):1164--1166.

\bibitem[Cochran et~al., 2004b]{Cochran2004}
Cochran, E.~S., Vidale, J.~E., and Tanaka, S. (2004b).
\newblock {Earth tides can trigger shallow thrust fault earthquakes}.
\newblock {\em Science}, 306(5699):1164--1166.

\bibitem[Craig et~al., 2017]{Craig2017}
Craig, T.~J., Chanard, K., and Calais, E. (2017).
\newblock {Hydrologically-driven crustal stresses and seismicity in the New
  Madrid Seismic Zone}.
\newblock {\em Nature Communications}, 8(1).

\bibitem[{Dalban Canassy} et~al., 2016]{DalbanCanassy2016}
{Dalban Canassy}, P., R{\"{o}}{\"{o}}sli, C., and Walter, F. (2016).
\newblock {Seasonal variations of glacier seismicity at the tongue of
  Rhonegletscher (Switzerland) with a focus on basal icequakes}.
\newblock {\em Journal of Glaciology}, 62(231):18--30.

\bibitem[Daley and Vere-Jones, 2003]{daley2003introduction}
Daley, D.~J. and Vere-Jones, D. (2003).
\newblock An introduction to the theory of point processes, volume 1:
  Elementary theory and methods.
\newblock {\em Verlag New York Berlin Heidelberg: Springer}.

\bibitem[Dunn, 1961]{Dunn1961}
Dunn, O.~J. (1961).
\newblock Multiple comparisons among means.
\newblock {\em Journal of the American Statistical Association},
  56(293):52--64.

\bibitem[Dutilleul et~al., 2015]{Dutilleul2015}
Dutilleul, P., Johnson, C.~W., Bürgmann, R., Wan, Y., and Shen, Z.-K. (2015).
\newblock Multifrequential periodogram analysis of earthquake occurrence: An
  alternative approach to the schuster spectrum, with two examples in central
  california.
\newblock {\em Journal of Geophysical Research: Solid Earth},
  120(12):8494--8515.

\bibitem[Eriksson et~al., 2009]{eriksson2009statistics}
Eriksson, J., Ollila, E., and Koivunen, V. (2009).
\newblock Statistics for complex random variables revisited.
\newblock In {\em 2009 IEEE International Conference on Acoustics, Speech and
  Signal Processing}, pages 3565--3568. IEEE.

\bibitem[Fisher, 1929]{fisher1929tests}
Fisher, R.~A. (1929).
\newblock Tests of significance in harmonic analysis.
\newblock {\em Proceedings of the Royal Society of London. Series A, Containing
  Papers of a Mathematical and Physical Character}, 125(796):54--59.

\bibitem[Heaton, 1975]{heaton1975tidal}
Heaton, T.~H. (1975).
\newblock Tidal triggering of earthquakes.
\newblock {\em Geophysical Journal International}, 43(2):307--326.

\bibitem[Heaton, 1982]{Heaton1982}
Heaton, T.~H. (1982).
\newblock {Tidal triggering of earthquakes}.
\newblock {\em Bull. Seis. Soc. Am.}, 72(6):2181--2200.

\bibitem[Heki, 2003]{Heki2003}
Heki, K. (2003).
\newblock {Snow load and seasonal variation of earthquake occurrence in Japan}.
\newblock {\em Earth and Planetary Science Letters}, 207(1-4):159--164.

\bibitem[Hernandez, 1999]{hernandez1999time}
Hernandez, G. (1999).
\newblock Time series, periodograms, and significance.
\newblock {\em Journal of Geophysical Research: Space Physics},
  104(A5):10355--10368.

\bibitem[Johnson et~al., 2017a]{Johnson2017}
Johnson, C.~W., Fu, Y., and B{\"{u}}rgmann, R. (2017a).
\newblock {Seasonal water storage, stress modulation, and California
  seismicity}.
\newblock {\em Science}, 356(6343):1161--1164.

\bibitem[Johnson et~al., 2017b]{Johnson2017a}
Johnson, C.~W., Fu, Y., and B{\"{u}}rgmann, R. (2017b).
\newblock {Stress Models of the Annual Hydrospheric, Atmospheric, Thermal, and
  Tidal Loading Cycles on California Faults: Perturbation of Background Stress
  and Changes in Seismicity}.
\newblock {\em Journal of Geophysical Research: Solid Earth},
  122(12):10,605--10,625.

\bibitem[Johnson et~al., 2020]{JOHNSON2020}
Johnson, C.~W., Fu, Y., and B{\"{u}}rgmann, R. (2020).
\newblock {Hydrospheric modulation of stress and seismicity on shallow faults
  in southern Alaska}.
\newblock {\em Earth and Planetary Science Letters}, 530:115904.

\bibitem[KNOTT, 1888]{knott1888earthquakes}
KNOTT, C.~G. (1888).
\newblock Earthquakes and earthquake sounds: As illustrations of the general
  theory of elastic vibrations.
\newblock {\em Transactions of the Seismological Society of Japan},
  (12):115--136.

\bibitem[Li and Zhan, 2018]{Li2018}
Li, Z. and Zhan, Z. (2018).
\newblock {Pushing the limit of earthquake detection with distributed acoustic
  sensing and template matching: A case study at the Brady geothermal field}.
\newblock {\em Geophysical Journal International}, 215(3):1583--1593.

\bibitem[M{\'{e}}tivier et~al., 2009]{Metivier2009}
M{\'{e}}tivier, L., de~Viron, O., Conrad, C.~P., Renault, S., Diament, M., and
  Patau, G. (2009).
\newblock {Evidence of earthquake triggering by the solid earth tides}.
\newblock {\em Earth and Planetary Science Letters}, 278(3-4):370--375.

\bibitem[Molchan and Dmitrieva, 1992]{Molchan1992a}
Molchan, G.~M. and Dmitrieva, O. (1992).
\newblock {Aftershock identification: methods and new approaches}.
\newblock {\em Geophys. J. Int}, 109:501--516.

\bibitem[Mu{\c{c}}o, 1999]{Muco1999}
Mu{\c{c}}o, B. (1999).
\newblock {Statistical investigation on possible seasonality of seismic
  activity and rainfall-induced earthquakes in Balkan area}.
\newblock {\em Physics of the Earth and Planetary Interiors},
  114(3-4):119--127.

\bibitem[Ng and Maechler, 2007]{Ng2007}
Ng, P. and Maechler, M. (2007).
\newblock A fast and efficient implementation of qualitatively constrained
  quantile smoothing splines.
\newblock {\em Statistical Modelling}, 7(4):315--328.

\bibitem[Ng and Maechler, 2017]{COBS}
Ng, P.~T. and Maechler, M. (2017).
\newblock {\em COBS -- Constrained B-splines (Sparse matrix based)}.
\newblock R package version 1.3-3.

\bibitem[Ogata, 1999]{Ogata1999}
Ogata, Y. (1999).
\newblock {Seismicity Analysis through Point-process Modeling: A Review}.
\newblock {\em Seismicity Patterns, their Statistical Significance and Physical
  Meaning}, 155:471--507.

\bibitem[Ogata, 2005]{Ogata2005}
Ogata, Y. (2005).
\newblock {Detection of anomalous seismicity as a stress change sensor}.
\newblock {\em Journal of Geophysical Research B: Solid Earth}, 110(5):1--14.

\bibitem[Omori, 1894]{Omori1894}
Omori, F. (1894).
\newblock {On the Aftershocks of Earthquake}.
\newblock {\em J. Coll. Sci. Univ. Tokyo}, 7:111--200.

\bibitem[Perol et~al., 2018]{Perol2018}
Perol, T., Gharbi, M., and Denolle, M. (2018).
\newblock {Convolutional neural network for earthquake detection and location}.
\newblock {\em Science Advances}, 4(2):2--10.

\bibitem[Rayleigh, 1880]{rayleigh1880xii}
Rayleigh, L. (1880).
\newblock Xii. on the resultant of a large number of vibrations of the same
  pitch and of arbitrary phase.
\newblock {\em The London, Edinburgh, and Dublin Philosophical Magazine and
  Journal of Science}, 10(60):73--78.

\bibitem[Reasenberg, 1985]{Reasenberg1985}
Reasenberg, P. (1985).
\newblock {Second-order moment of central California seismicity, 1969–1982}.
\newblock {\em Journal of Geophysical Research}, 90(B7):5479.

\bibitem[Ross et~al., 2019]{Ross2019}
Ross, Z.~E., Trugman, D., Hauksson, E., and Shearer, P. (2019).
\newblock {Searching for hidden earthquakes in Southern California}.
\newblock {\em Science}, 6888(April).

\bibitem[Rydelek and Hass, 1994]{rydelek1994estimating}
Rydelek, P.~A. and Hass, L. (1994).
\newblock On estimating the amount of blasts in seismic catalogs with
  schuster's method.
\newblock {\em Bulletin of the Seismological Society of America},
  84(4):1256--1259.

\bibitem[Schuster, 1897]{schuster1897lunar}
Schuster, A. (1897).
\newblock On lunar and solar periodicities of earthquakes.
\newblock {\em Proceedings of the Royal Society of London},
  61(369-377):455--465.

\bibitem[Shudde and Barr, 1977]{Shudde1977}
Shudde, R. and Barr, D. (1977).
\newblock {An analysis of earthquake frequency data}.
\newblock {\em Bull. Seis. Soc. Am.}, 67(5):1379--1386.

\bibitem[Simpson et~al., 1988]{Simpson1988}
Simpson, D., Leith, W., and Scholz, C. (1988).
\newblock {Two types of reservoir-induced seismicity}.
\newblock {\em Bull. Seis. Soc. Am.}, 78(6):2025--2040.

\bibitem[Smirnov et~al., 2018]{Smirnov2018}
Smirnov, V.~B., Mikhailov, V.~O., Ponomarev, A.~V., Arora, K., Chadha, R.~K.,
  Srinagesh, D., and Potanina, M.~G. (2018).
\newblock {On the Dynamics of the Seasonal Components of Induced Seismicity in
  the Koyna–Warna Region, Western India}.
\newblock {\em Izvestiya, Physics of the Solid Earth}, 54(4):632--640.

\bibitem[Stroup et~al., 2007]{Stroup2007}
Stroup, D.~F., Bohnenstiehl, D.~R., Tolstoy, M., Waldhauser, F., and Weekly,
  R.~T. (2007).
\newblock {Pulse of the seafloor: Tidal triggering of microearthquakes at 9
  degrees 50 minutes N East Pacific Rise}.
\newblock {\em Geophysical Research Letters}, 34(15):1--6.

\bibitem[Talwani, 1997]{Talwani1997}
Talwani, P. (1997).
\newblock {On the Nature of Reservoir-induced Seismicity}.
\newblock In S., T., editor, {\em Seismicity Associated with Mines, Reservoirs
  and Fluid Injections}. Birkh{\"{a}}user, Basel, pageoph to edition.

\bibitem[Tanaka, 2012]{Tanaka2012}
Tanaka, S. (2012).
\newblock {Tidal triggering of earthquakes prior to the 2011 Tohoku-Oki
  earthquake (Mw 9.1)}.
\newblock {\em Geophysical Research Letters}, 39(7):1--4.

\bibitem[Tanaka et~al., 2002a]{tanaka2002evidence}
Tanaka, S., Ohtake, M., and Sato, H. (2002a).
\newblock Evidence for tidal triggering of earthquakes as revealed from
  statistical analysis of global data.
\newblock {\em Journal of Geophysical Research: Solid Earth}, 107(B10):ESE--1.

\bibitem[Tanaka et~al., 2002b]{Tanaka2002}
Tanaka, S., Ohtake, M., and Sato, H. (2002b).
\newblock {Evidence for tidal triggering of earthquakes as revealed from
  statistical analysis of global data}.
\newblock {\em Journal of Geophysical Research: Solid Earth}, 107(B10):ESE
  1--1--ESE 1--11.

\bibitem[Tanaka et~al., 2004]{Tanaka2004}
Tanaka, S., Ohtake, M., and Sato, H. (2004).
\newblock {Tidal triggering of earthquakes in Japan related to the regional
  tectonic stress}.
\newblock {\em Earth, Planets and Space}, 56(5):511--515.

\bibitem[Tolstoy et~al., 2002]{Tolstoy2002}
Tolstoy, M., Vernon, F.~L., Orcutt, J.~A., and Wyatt, F.~K. (2002).
\newblock {Breathing of the seafloor: Tidal correlations of seismicity at Axial
  volcano}.
\newblock {\em Geology}, 30(6):503--506.

\bibitem[Tsuruoka et~al., 1995]{Tsuruoka1995}
Tsuruoka, H., Ohtake, M., and Sato, H. (1995).
\newblock {Statistical test of the tidal triggering of earthquakes:
  contribution of the ocean tide loading effect}.
\newblock {\em Geophysical Journal International}, 122(1):183--194.

\bibitem[Ueda and Kato, 2019]{Ueda2019}
Ueda, T. and Kato, A. (2019).
\newblock {Seasonal Variations in Crustal Seismicity in San-in District,
  Southwest Japan}.
\newblock {\em Geophysical Research Letters}, 46(6):3172--3179.

\bibitem[Utsu and Ogata, 1995]{Utsu1995}
Utsu, T. and Ogata, Y. (1995).
\newblock {The centenary of the omori formula for a decay law of aftershock
  activity}.
\newblock {\em Journal of Physics of the Earth}, 43(1):1--33.

\bibitem[Utsu et~al., 1995]{utsu1995centenary}
Utsu, T., Ogata, Y., et~al. (1995).
\newblock The centenary of the omori formula for a decay law of aftershock
  activity.
\newblock {\em Journal of Physics of the Earth}, 43(1):1--33.

\bibitem[Vidale et~al., 1998]{Vidale1998}
Vidale, J.~E., Agnew, D.~C., Johnston, M. J.~S., and Oppenheimer, D.~H. (1998).
\newblock {Absence of earthquake correlation with Earth tides: An indication of
  high preseismic fault stress rate}.
\newblock {\em Journal of Geophysical Research: Solid Earth},
  103(B10):24567--24572.

\bibitem[Wang and Shearer, 2015]{Wang2015}
Wang, W. and Shearer, P.~M. (2015).
\newblock {No clear evidence for localized tidal periodicities in earthquakes
  in the central Japan region}.
\newblock {\em Journal of Geophysical Research: Solid Earth},
  120(9):6317--6328.

\bibitem[Wilcock, 2009]{Wilcock2009}
Wilcock, W. S.~D. (2009).
\newblock {Tidal triggering of earthquakes in the northeast Pacific Ocean}.
\newblock {\em Geophys. J. Int.}, 179:1055--1070.

\bibitem[Wilcox, 2001]{Wilcox2001}
Wilcox, W. (2001).
\newblock {Tidal triggering of microearthquakes on the Juan de Fuca Ridge}.
\newblock {\em Geophys. Res. Lett.}, 28(20):3999--4002.

\bibitem[Yoon et~al., 2015]{Yoon2015}
Yoon, C.~E., O'Reilly, O., Bergen, K.~J., and Beroza, G.~C. (2015).
\newblock {Earthquake detection through computationally efficient similarity
  search}.
\newblock {\em Science Advances}, 1(11):1--14.

\bibitem[Zaliapin et~al., 2008]{Zaliapin2008}
Zaliapin, I., Gabrielov, A., Keilis-Borok, V., and Wong, H. (2008).
\newblock {Clustering Analysis of Seismicity and Aftershock Identification}.
\newblock {\em Physical Review Letters}, 101(1):018501.

\bibitem[Zhuang et~al., 2002]{Zhuang2002}
Zhuang, J., Ogata, Y., and Vere-Jones, D. (2002).
\newblock {Stochastic declustering of space-time earthquake occurrences}.
\newblock {\em Journal of the American Statistical Association},
  97(458):369--380.

\end{thebibliography}

\appendix

\section{Order Statistic Property of Poisson Processes}
We introduce a well-known result in treatment of events from a Poisson process, and concomitant notation of which we will make use extensively.

\begin{Thm}\label{Thm:osp}
The property of being a samples from a Poisson processes, and being the order statistics of an i.i.d. sample, are equivalent in the following sense.\\
As the first part of this theorem, the following two statements are true:
\begin{enumerate}
\itemsep-0.2em
\item[(i)] Consider a Poisson process over a bounded real interval $I\subseteq\RR,$ with Poisson distributed observation count variable $N$ and $\tau=(T_1,\dots, T_N)$ a sample. Then, there exists an random variable $X$, absolutely continuous over $I$, and i.i.d.copies $X_1,X_2,\dots$ of $X$, such that $(X_1,\dots, X_N)$ is identical in distribution to $\left(T_{S(1)},\dots, T_{S(N)}\right)$, where the conditional $S|N$ is a uniform random permutation of the index set $[N]$.
\item[(ii)] Let $X$ be a random variable, absolutely continuous over a compact real interval $I\subseteq\RR,$ let $N$ be Poisson distributed. Let $X_1,\dots, X_N$ be copies of $X$, conditionally i.i.d., on $N$. Let $T_i:=X_{(i)}$ the $i$-th order statistic of the sample $(X_1,\dots, X_N)$, ties resolved uniformly. Then, the sample $\tau:=(T_1,\dots, T_N)$ is a sample from a Poisson process over $I$.
\end{enumerate}
As the second part of the theorem, consider the collections of objects as assumed (in premises) and implied (by consequences) in the statements (i) and (ii).\\
Then, these two collections of objects are canonically isomorphic, i.e., concordant choices for premise symbols will yield the same choices for constructed consequence symbols (with ``same'' meaning ``isomorphic as collections of random variables'').\\
Furthermore, the equality $\lambda = \nu \cdot p_{X|N}$ holds between the pdf $p_{X|N}$ of $X|N$, the rate constant $\nu$ of $N$, and the intensity function $\lambda$ of $\tau$.
\end{Thm}
\begin{proof}
This is directly implied by discussion in Section 7.1, in particular Example 7.1(a), of~\cite{daley2003introduction}, which phrases the matter in terms of laws of random variables, without introducing notation for the random variables themselves. The correspondence between our notation and notation ibidem is: our $N$ is $n$ ibidem; our $I$ is $A$ ibidem; our joint conditional law of $X_1,\dots,X_N|N$ (no symbol introduced here) is $j_n$ ibidem;  our $p_{X|N}$ is $\phi$ ibidem; our $\nu$ is $C$ ibidem.\\
Additional, simplifying assumptions for the situation ibidem are implied by the statement of this theorem: we assume that the domain is univariate, i.e., $d=1$ ibidem. In addition, we assume that $N$ is Poisson, implying finiteness of the measure $\lambda$ on $A$ ibidem.\\
The statements in this theorem are then implied by definitions, and the canonicity of the objects established ibidem.
\end{proof}

Theorem~\ref{Thm:osp} is sometimes referred to as ``Poisson process samples have the order statistics property'', though this terminology is slightly inexact in its lack of reference to the random size of the sample (the random $N$ in the theorem must be Poisson).

The original publications around the Schuster spectrum test~\cite{Ader2013} also make use of this correspondence implicitly by representing the process $\tau$ by the pdf of $X$ - which may be somewhat confusing for a reader unfamiliar with Theorem~\ref{Thm:osp}, but is easily recognized by one who is.

However, as our situation is more general, with primary and secondary shocks, where the secondary shocks are only conditionally Poisson, we need to introduce more stringent notation as identification and conceptualization is no longer obvious.

\section{Basic results on complex random variables}
\label{sec:complexRV}

The proofs of our main results heavily make use of complex (number valued) random variables which arise in Fourier representations of signals with a suspected periodicity.

The results we use are natural generalizations of known results on expectation and variance of real random variables (univariate and multivariate); for complex random variable, many are straightforward generalizations which occasionally appear implicitly in literature when needed. However, we were unable to find a citable reference for those results we use (the reference~\cite{eriksson2009statistics} has some), therefore we collect and present necessary, ancillary, and related statements below.

Hence, proofs are provided below.
\begin{Def}\label{Def:complexrv}
A random variable, taking values in $\CC$, is any random variable $X$ which can be written as $X=Y+\imath \cdot Z$, where $Y,Z$ are random variables taking values in $\RR$, and $\imath$ is the complex unit.\\
In the above case, we write $Y:=\Real X$, and $Z:=\Imag X$, calling $\Real X$ the \emph{real part} of $X$, and $\Imag X$ the \emph{imaginary part} of X.
\end{Def}
Thus, formally, complex random variables are identified with $\RR^2$-valued random variables, via concatenation with the canonical bijection $\CC\cong \RR^2,\; x\mapsto (\Real x, \Imag x)$ which identifies each complex number uniquely with its point in the ``complex plane''.\\
We identify real variables with complex random variables whose imaginary part is zero with probability one, and carry through the above identification for conditional random variables.

\begin{Def}\label{Def:complexdefs}
Let $X$ be a random variable, taking values in $\CC$.
\begin{enumerate}
\itemsep-0.2em
\item[(i)] The \emph{conjugate} of $X$ is defined as $X^\ast:= \Real X - \imath \cdot \Imag X$.
\item[(ii)] The \emph{modulus} of $X$ is defined as $|X| := \sqrt{(\Real X)^2 + (\Imag X)^2} = \sqrt{X\cdot X^\ast}$.
\item[(iii)] The \emph{expectation} of $X$ is defined as $\EE[X]:=\EE[\Real X]+ \imath\cdot \EE[\Imag X]$.
\item[(iv)] The (complex) \emph{variance} of $X$ is defined as
        $\Var [X] := \EE[X\cdot X^\ast] - \EE[X]\cdot \EE[X]^\ast$.
\end{enumerate}
\end{Def}

It is very important to note that the (complex) variance $\Var [X]$ is not the same as the variance of the bivariate real random variable $(\Real X, \Imag X)$, which is a $(2\time 2)$ covariance matrix. However, they are closely related:

\begin{Lem}\label{Lem:complexvar}
Let $X$ be a random variable, taking values in $\CC$.\\
The following numbers (possibly infinite) are the same:
\begin{enumerate}
\itemsep-0.2em
\item[(i)] $\Var [X]$
\item[(ii)] $\EE\left[|X|^2\right]-\left|\EE[X]\right|^2$
\item[(iii)] $\EE\left[|X-\EE[X]|^2\right]$
\item[(iv)] $\frac{1}{2}\EE\left[|X-X'|^2\right],$ where $X'$ is any independent copy of $X$
\item[(v)] $\Var[\Real X] + \Var [\Imag X]$
\end{enumerate}
\end{Lem}
\begin{proof}
This all follows by elementary computation, expanding to real random variables and then applying rote calculation laws there (e.g., linearity of expectation).
\end{proof}

Lemma~\ref{Lem:complexvar}~(v) immediately connects complex variance to properties and results about real variance:

\begin{Lem}
\label{Lem:complexvarprops}
Let $X,X_1,\dots, X_N$ be random variables, taking values in $\CC$. Then,
\begin{enumerate}
\itemsep-0.2em
\item[(i)] $\Var [X] \ge 0$
\item[(ii)] If $X_1,\dots, X_N$ are mutually independent, then
$$\Var \left[\sum_{i=1}^N X_i\right] = \sum_{i=1}^N \Var \left[X_i\right]$$
\end{enumerate}
\end{Lem}
\begin{proof}
This is a direct consequence of Lemma~\ref{Lem:complexvar}~(v) and well-known properties of the variance: for (i), that $\Var [Z] \ge 0$ for any real random variable; for (ii), that $\Var \left[\sum_{i=1}^N Z_i\right] = \sum_{i=1}^N \Var \left[Z_i\right]$ for mutually independent real random variables $Z_1,\dots, Z_N$.
\end{proof}

We further prove some useful results about conditionals expectations and varianes:

\begin{Lem}\label{Lem:complexpect}
Let $X$ be a complex random variable, let $Y$ be any random variable such that the conditional $X|Y$ is defined. Then:
\begin{enumerate}
\itemsep-0.2em
\item[(i)] $\EE [X] = \EE\left[\EE[X|Y]\right]$
\item[(ii)] $\Var [X] = \EE\left[\Var[X|Y]\right] + \Var \left[\EE[X|Y]\right]$.
\end{enumerate}
when both sides of the respective equation are finite and well-defined.
\end{Lem}
\begin{proof}
This all follows by elementary computation, expanding to real random variables and then applying rote calculation laws there (e.g., linearity of expectation).\\
Very direct proofs are obtained as follows:\\
For (i), substitute the definition of $\EE$ on both sides, use linearity of (conditional) expectation, apply the law of total expectation for real random variables.\\
For (ii), proceed as follows: for all variances of complex numbers, use equivalence between (i) and (v) in Lemma~\ref{Lem:complexvar}; for all expectations of complex numbers, use the definition. Then, apply the law of total variance (``EVE law'') for real random variables.
\end{proof}

It should be pointed out that while Lemma~\ref{Lem:complexpect} is an easy consequence, it is not an immediate corollary of the real law of total variance: both variances on the RHS are complex variances, and the first expectation is real, while the second is complex.

\section{Phasors, characteristic functions, Fourier transforms, and phase invariance}

In our proof, we will also make heavy use of random phases and phasors to encode periodicities. The cumulants of these, in turn, are known to be closely related to characteristic functions and Fourier transforms of probability densities. We introduce notation and highlight key correspondences in this section, below.

\begin{Def}\label{Def:phasor}
Let $a$ and $\theta$ be real numbers or random variables taking values in the reals. We will write $a\angle \theta := a\cdot \exp(\imath\cdot \theta)$, where $\imath$ is the complex unit. The (possibly random) quantity $a\angle \theta$ is called \emph{phasor} with \emph{amplitude} $a$ and \emph{phase} $\theta$ (especially in engineering and physics literature).\\
We also abbreviate $1\angle \theta$ by $\angle \theta$.
\end{Def}

Advantages of phasor notation are compatibility with modulus and multiplication, i.e., $|a\angle \theta| = |a|$ and $(a\angle \theta)\cdot (b\angle \theta') = (ab)\angle(\theta + \theta')$, calculation rules that are used below.

\begin{Def}\label{Def:charfun}
Let $X$ be a random variable taking values in $\RR$. The \emph{characteristic function} (cf) of $X$ is defined as $\chi_X: S\rightarrow \CC\;;\; t\mapsto \EE[\angle (tX)],$ where $S$ is set of $t$ where $\EE[\angle (tX)]$ is well-defined. $S$ is called the \emph{domain of convergence} of $\chi_X$.
\end{Def}

As well-known stylized results, it always holds $\RR\subseteq S$. Furthermore, $\chi_X$ is distribution defining for all $X$ with domain of convergence being all of $\CC$. That is, the law of any such $X$ is uniquely determined by its characteristic function over $\RR$. This statement becomes false when considering $\chi_X$ with arbitrary domain of convergence.

\begin{Def}\label{Def:Fourier}
Let $p:\RR\rightarrow \RR$ be an integrable function over $\RR$, i.e, $\RR\in L^1(\RR)$. The \emph{Fourier transform} $\calF p$ of $p$ is defined as
$$\calF p: \RR\rightarrow \RR\;;\; y\mapsto \int_\RR f(x) \angle (- 2\pi xy) \diff x.$$
Notationally, the operator $\calF$ applies before point-wise evaluation, i.e., we will write $\calF p(y)$ instead of $(\calF p)(y)$.
\end{Def}

Well-known stylized results from Fourier theory assert that $\calF p \in L^1(\RR)$ as well, and $\calF$ is unitary (as long as the normalization convention in Definition~\ref{Def:Fourier} is used). It should also be noted that the pdf of (univariate) real random variables are automatically in $L^1(\RR)$, hence always possesses a Fourier transform.

\begin{Lem}\label{Lem:Fourier}
Let $X$ be an absolutely continuous (w.r.t. Lebesgue measure) real random variable, with pdf $p_X$ and cf $\chi_X$.
The following equalities hold for any $t\in \RR$:
\begin{enumerate}
\itemsep-0.2em
\item[(i)] $\EE [\angle (tX)] = \EE \left[(\angle X)^t\right] = \calF p_X\left(-\frac{t}{2\pi}\right) = \chi_X(t)$
\item[(ii)] $\Var [\angle (tX)] = 1 - |\chi_X(t)|^2 = 1 - \left|\calF p_X \left(-\frac{t}{2\pi}\right)\right|^2.$
\end{enumerate}
\end{Lem}
\begin{proof}
This is immediate from simple elementary computation, following substitution of definitions.
\end{proof}

Some basic properties of random phasors' cumulants, most well-known from the closely related context of distributions on the circle:
\begin{Prop}\label{Prop:phasor}
Let $\theta$ be a random variable, taking values in $\RR$. Let $X:= \angle \theta$. Then:
\begin{enumerate}
\itemsep-0.2em
\item[(i)] $\Var [X] = 1 - \left|\EE[X]\right|^2$
\item[(ii)] $0\le \left|\EE[X]\right| \le 1$ and $0\le \left|\Var[X]\right| \le 1$
\item[(iii)] $|\EE[X]| = 1$ iff $\Var[X] = 0$ iff $X$ is a constant random variable
\item[(iv)] $\Var[X] + \left|\EE[X] - \angle \EE[\theta]\right|^2 = \EE\left[|X-\angle \EE[\theta]|^2\right]\le \Var [\theta]$
\item[(v)] In particular, $\Var [X] \le \Var[\theta]$
\end{enumerate}
\end{Prop}
\begin{proof}
For the proofs, write $S:=\{x\in \CC\;:\; |x| = 1\}$ for the unit circle in the complex plane. Note that $S$ is the image of the map $\RR\rightarrow S,\; \alpha\mapsto \angle \alpha$.\\
(i) is the equivalence of (i) and (ii) in Lemma~\ref{Lem:complexvar}, observing that $|X^2| = |\angle \theta|^2 = 1$.\\
(ii) $0\le \left|\Var[X]\right|$ follows from the equivalence of (i) and (v) in Lemma~\ref{Lem:complexvar}, and non-negativity of (real) variance. $0\le \left|\EE[X]\right|$ is because $|.|$ is a norm. The remaining inequalities follow from these inequalities, the equality in (i), and elementary computation.\\
(iii) $|\EE[X]| = 1$ iff $\Var[X] = 0$ is implied by (i). Equivalence of (i) and (v) in Lemma~\ref{Lem:complexvar} implies that $\Var[X] = 0$ iff $\Var[\Real X] = \Var[\Imag X] = 0$. The latter, by basic theory of (real) variance, holds iff both $\Real X$ and $\Imag X$ are constant random variables, which holds iff $X$ is a constant random variable.\\
(iv) The first equality is the complex analogue of the bias-variance decomposition of the mean squared error, which one verifies just as in the real case by elementary computation.\\
For the inequality, note that $\Var [\theta] = \EE[(\theta - \EE[\theta])^2]$. Identifying $S$, the image of $\angle (.),$ with the unit circle in the complex plane (as above in the pre-amble), we observe that $\left|X-\angle \EE[\theta]\right| = \left|\angle \theta -\angle \EE[\theta]\right|\le \left|\theta - \EE[\theta]\right|$ is implied by the schoolbook theorem ``chord length is less or equal arc length'' for any circle (and any arc connecting the chord, irrespective of orientation and multiplicity). Taking squares and expectations yields the claim.\\
(v) This follows directly from (iv), from comparing leftmost and rightmost side, and observing that $\left|\EE[X] - \angle \EE[\theta]\right|\ge 0$ because $|.|$ is a norm.

\end{proof}

Angular invariance defines an important class of distributions:

\begin{Def}
A complex random variable $Z$ is called circular symmetric if $Z$ and $Z\cdot \angle \theta$ are identically distributed for any $\theta\in \RR$.
\end{Def}

\begin{Prop}\label{Prop:circrv}
Let $Z$ be a circular symmetric complex random variable.  Then,
\begin{enumerate}
\itemsep-0.2em
\item[(i)] $Z$ and $Z\cdot \angle \theta$ are identically distributed for any real random variable $\theta$.
\item[(ii)] $Z\cdot Y$ is circular symmetric for any complex random variable $Y$.
\item[(iii)] If $Y$ is circular symmetric, so is $Y+Z$.
\item[(iv)] If $\EE[Z]$ exists, then $\EE[Z] = 0$.
\item[(v)] Let $X:=(\Real Z, \Imag Z)$ be the bivariate obtained by identification of the domain of $Z$ with $\RR^2$ (``the complex plane''). Then, $\Var[Z] = \frac{\Var[X]}{2} \cdot I$, with $I\in \RR^{2\times 2}$ being the bivariate identity matrix.
\end{enumerate}
\end{Prop}
\begin{proof}
(i) By assumption of circular symmetry, it holds that the conditional distribution $Z\cdot \angle \theta |\theta$ is distributed according to $Z$, and does not depend on the specific value of $\theta$. Therefore, the unconditional distributions $Z\cdot \angle \theta$ and $Z$ must be identically distributed.\\
(ii) By assumption of circular symmetry, $(\angle\theta)\cdot Z$ and $Z$ are identically distributed for arbitrary $\theta\in \RR$. Hence, $((\angle \theta)\cdot Z)\cdot Y$ and $Z\cdot Y$ are identically distributed. Since $\theta$ was arbitrary, $Z\cdot Y$ is circular symmetric.\\
(iii)By assumption of circular symmetry, $(\angle\theta)\cdot Z$ and $Z$ are identically distributed, and $(\angle\theta)\cdot Y$ and $Y$ are identically distributed, for arbitrary $\theta\in \RR$. Hence, $(\angle \theta)\cdot (Z + Y)$ and $Z+Y$ are identically distributed. Since $\theta$ was arbitrary, $Z+Y$ is circular symmetric.\\
(iv) By the assumption of circular symmetry, $\EE[Z] = \EE[Z]\cdot \angle x$ for any $x\in \RR$. Therefore, $\arg \EE[X] = \arg\EE[X] + x \mod 2\pi $ for any $x\in \RR$, if $\arg \EE[X]$ exists. In particular, $0 = x\mod 2\pi$ for any $x$ if $\arg \EE[X]$ exists, therefore $\arg \EE[X]$ does not exist, and therefore $\EE[X] =0$. The claim on $\EE[Z]$ follows from definition of $Z$.\\
(v) From Proposition~\ref{Lem:complexvar}~(v), $\Var[X] = \Tr\Var[Z]$. Circular invariance implies that $Z$ and $Z' := \imath Z = \angle{\pi/2}\cdot Z$ have the same distribution, thus $X' = (\Real Z', \Imag Z') = (\Imag Z, -\Real Z)$ and $X$ must have the same distribution. Thus, $\Var[X'] = \Var[X]$, which implies $\Var(\Imag Z) = \Var(\Real Z)$, and $\Cov (\Real Z, \Imag Z) = - \Cov (\Real Z, \Imag Z)$. The latter implies $\Cov (\Real Z, \Imag Z) = 0$, the former, together with $\Var[X] = \Tr\Var[Z]$, implies $\Var[X] = 2\Var(\Real Z) = 2\Var(\Imag Z)$. Altogether, these imply the claim.
\end{proof}

An important special case is the uniform on the unit circle:

\begin{Lem}\label{Lem:unifcirc}
Let $\theta\sim\Unif([0,2\pi])$. Let $X:= \angle \theta$, let $Z:=(\Real X, \Imag X)$ be the bivariate obtained by identification of the domain of $X$ with $\RR^2$ (``the complex plane''). Then:
\begin{enumerate}
\itemsep-0.2em
\item[(i)] $X\sim \Unif(\{x\in \CC\;:\;|x|=1\}),$ and $Z\sim \Unif(\{x\in \RR^2\;:\;|x|_2=1\})$
\item[(ii)] $X$ is circular symmetric.
\item[(iii)] $\EE[X] = 0,$ and $\EE[Z] = (0,0)$
\item[(iv)] $\Var[X] = 1,$ and $\Var[Z] = \frac{1}{2} I$, with $I\in \RR^{2\times 2}$ being the bivariate identity matrix
\end{enumerate}
\end{Lem}
\begin{proof}
(i) this is a direct consequence of the transformation theorem for multi-variate random variables (in this case: univariate to bivariate).\\
(ii) This follows from rotation invariance of the uniform distribution on the circle (as a possible defining property). \\
(iii) This is a consequence of Proposition~\ref{Prop:circrv}~(iii).The claim on $\EE[Z]$ follows from definition of $Z$.\\
(iv) By (iii) and definition of complex variance, $\Var[X] = \EE[|X|^2]$. By (ii), $|X|=1$ almost surely, thus $\Var[X] = 1$. The statement for $Z$ is a consequence of Proposition~\ref{Prop:circrv}~(v).
\end{proof}

\section{The Schuster test}
\label{sec:app-schuster}
We present the ``Schuster test'' (also referred to as ``Schuster/Fischer procedure''), in a consensus form, based on implicit use in literature when this term is referred to, in the geoscience domain.

\subsection{Generative setting}
\label{sec:app-schuster.gen}
The following generative setting is usually assumed for a Schuster test:
$(T_n)_{n \in 1, \ldots, N}$ with $N\sim \Poi(\nu_0),$ assumed a sample from a (possibly heterogeneous) Poisson process, modelling the events.\\
By Theorem~\ref{Thm:osp}, it is no loss of generality (by identification with the i.i.d.~sample presentation) to also assume the following:
The sample $T_1,\dots,T_N$ is an i.i.d.~sample (conditional on $N$).

\subsection{The Schuster distance}
\label{sec:app-schuster.distasy}

The Schuster distance is defined as
$$D(\omega) := \sum_{i=1}^N \angle (\omega T_i).$$
Intuitively, this is simply the discrete Fourier transform of the event sequence, i.e., a constant signal, but sampled at time points which in general are not from a grid. Testing for periodicity is modelled as testing for a peak, for some specific frequency $\omega$ - as the signal is constant, the peak must be caused by the sampling process.

As the $T_i$ are i.i.d.~conditional on $N$ (as one can assume according to discussion in Section~\ref{sec:app-schuster.gen}), so are $\angle (\omega T_i), 1\le i\le N$, and so are the pairs of real random variables $Z_i:=(\Real \angle (\omega T_i), \Imag \angle (\omega T_i)), 1\le i\le N$, say, $Z_1,\dots, Z_N\sim Z$ for some generative $Z$. The common argumentation proceeds by conditioning on $N$, i.e., considering the sample size fixed.

Writing $X:= \Real D(\omega), Y:= \Imag D(\omega)$, according to the multivariate central limit theorem (note that $Z_i$ are compactly supported and therefore satisfy assumptions), one has $\left((X, Y) - \EE[Z]\right)/\sqrt{N} \approx \calN (0, \Var(Z))$, subject to the usual quantitative bounds on the approximation (e.g., Berry-Esseen, and the heuristic that the approximation is ``good'' if $N\ge 30$).

Under the null hypothesis, the Poisson process from which $T_n$ are drawn is assumed homogenous, thus $\angle T_i \sim \Unif \{x\in \CC\;:\;|x|=1\}$, and $Z\sim \Unif \{x\in \RR^2\;:\;|x|_2=1\}$. An elementary calculation yields that $\EE[Z] = (0,0)$, and $\Var(Z) = \frac{1}{2} I,$ with $I\in \RR^{2\times 2}$ the identity matrix. Therefore, under the null, $\left(X/\sqrt{N}, Y/\sqrt{N}\right)\approx \calN (0, \frac{1}{2} I).$

\subsection{The squared Schuster distance and the Schuster p-statistic}
\label{sec:app-schuster-test}

The \emph{squared Schuster distance} is defined as $D^2(\omega):= |D(\omega)|^2 = D(\omega)D(\omega)^\ast$.\\
Using the above notation, an elementary calculation shows that $D^2(\omega)/N = (X^2 + Y^2)/N  =(X/\sqrt{N})^2 + (X/\sqrt{N})^2.$
Continuing under assumption of the null: as $\left(X/\sqrt{N}, Y/\sqrt{N}\right)\approx \calN (0, \frac{1}{2} I),$ (by virtue of continuous mapping theorems) it also holds that $D^2(\omega)/N\approx \Exp(1)$ (in the sense of convergence in distribution with quantifiable error bounds that are small for all practical purposes if $N\ge 30$) by virtue of the following well-known result:

\begin{Lem}\label{Lem:normexp}
Let $(U,V)\sim \calN (0,a\dot I)$ with $a\in \RR_+,$ with $I\in \RR^{2\times 2}$ the identity matrix and $0$ the zero of $\RR^2$.\\
Then, $U^2+V^2\sim \Exp \left(2 a\right)$.
\end{Lem}
\begin{proof}
This is an elementary calculation, using standard results on multivariate transformation of random variables.
\end{proof}

On a historical note, that $D^2(\omega)/N\approx \Exp(1)$ (under the null) was probably first derived in~\cite{rayleigh1880xii}. The connection to the null of a Schuster-like test is made by Schuster~\cite[paragraph 2]{schuster1897lunar}.

To obtain the Schuster test in its contemporary form, as a frequentist hypothesis test, we use another well-known result to derive the p-value:

\begin{Lem}
Let $U\sim \Exp(1)$. Then $\exp (- U)\sim \Unif [0,1]$.
\end{Lem}
\begin{proof}
This is also an elementary calculation, using standard results on univariate transformation of random variables.
\end{proof}

to infer that
$p:=\exp\left(-D^2(\omega)/N\right)\approx \Unif [0,1].$

The above chain of arguments shows that $p\le \alpha$ iff $F_P(|D(\omega)|)\ge 1-\alpha$, for the random variable $P\sim \Exp (N),$ where $|D(\omega)| \approx P$ is a ``good'' approximation (if $N\ge 30$).

Therefore, $p$ is an approximate $p$-value for $|D(\omega)|$ taking uncharacteristically large values, with a rejection region of $D(\omega)$ being in $\{x\in \CC\;:\; |x|\ge \sqrt{N}\}.$

It should be noted that the above properties are completely independent of the particular value of $\omega$, except through the rejection region which changes with $\omega$, and thus gives a \emph{different test}, for each choice of
$\omega$.

\section{The Schuster Test subject to an Aftershock Process}

This section contains derivation of the main results of the manuscript, on the Schuster spectrum under an afterschock process.

\subsection{Generative setting}
\label{sec:app-proof.setting}
We will assume the following generative setting (note that we re-define some symbols used in Section~\ref{sec:app-schuster.gen} in the new scope of this section):
$\tau:=(\tau_m)_{m \in 1, \ldots, M}$ with $M\sim \Poi(\nu_0),$ assumed a sample from a Poisson process, modelling the primary shocks.\\
$\varsigma_m:=(\varsigma_{mj}\})_{j \in 1, \ldots, A_m}$ with $A_m\sim \Poi(\nu),$ for $m=1,\dots, M$. Conditional on $\tau_m$, it is assumed that $\varsigma_m$ is a sample from a Poisson process, modelling secondary shocks.
It is further assumed that $\varphi_{mj}:=\varsigma_{mj}-\tau_m$ are identically distributed, independent of $\tau$, and with non-negative support (i.e., aftershocks happen after their respective primary shock.

By Theorem~\ref{Thm:osp}, it is no loss of generality (by identification with the i.i.d.~sample presentation) to also assume the following:
The sample $\tau_1,\dots,\tau_M$ is an i.i.d.~sample, with characteristic function $\chi_\tau$ (conditional on $M$).
The (pooled) sample $\varphi_{mj}, j = 1, \dots, A_m, m = 1,\dots, M$, is an i.i.d.~sample, independent of $\tau$ (and all $A_m$), with characteristic function $\chi_\varphi$. The pdf $g$ has non-negative support by the assumption on $\varphi_{mj}$ above.
We would also like to note that, according to Theorem~\ref{Thm:osp}, the densities and hence characteristic functions $\chi_\tau$ and $\chi_\varsigma$ do \emph{not} depend on the values of $M$ or $A_m$.

We further assume that the times of the shocks are known to the observer, but it is not necessarily known to the observer which shocks are primary shocks, or aftershocks. For this, we denote by $T_1,\dots, T_N$, with $N= M + \sum_{i=1}^M A_i$ the sample of all time points $\tau_m,\varsigma_{mj}$, in order of their occurrence, i.e., the order statistics of the pooled samples $\tau,\varsigma_1,\dots,\varsigma_M$. Practically and algorithmically usable statements can make reference to this sample only - while theoretical considerations may make reference to the unobserved knowledge of which events are primary shocks, which are secondary shocks, and which secondary shocks are associated with which primary shock.

We note that our setting is equivalent to that of a Hawkes process where aftershocks cannot have aftershocks, or of a re-parameterized Hawkes process.

\subsection{Angle representation of Schuster distance}
\label{sec:app-proof.SchusterDsymbols}

The Schuster distance is defined as
$$D(\omega) := \sum_{i=1}^N \angle (\omega T_i) = \sum_{m=1}^M \angle (\omega \tau_i) + \sum_{j=1}^{A_m} \angle (\omega \varsigma_{mj}),$$
where $\omega$ is some (radian) frequency value. Below, we also use the period length parameterization $\omega =: \frac{2\pi}{k}$, and will write $\tilde{T}_i:=\omega\cdot T_i, \tilde{\tau}_i := \omega\cdot \tau_i$, and so on, for products with $\omega$.\\
Note that the first expression of $D(\omega)$, in terms of $T_i$ is algorithmically computable under the assumptions, while the second in terms of $\tau_i$ and $\varsigma_{mj}$ is not (as our observer does not know which shocks are primary resp.~aftershocks). We also make note of two useful equalities:
$$D(\omega) = \sum_{m=1}^N \angle \tilde{T}_i = \sum_{m=1}^M \angle \tilde{\tau}_m\cdot\left( 1 + \sum_{j=1}^{A_m} \angle \tilde{\varphi}_{mj}\right).$$
Writing $S_i:= \angle \tilde{\tau}_i\cdot\left( 1 + \sum_{j=1}^{A_i} \angle \tilde{\varphi}_{ij}\right)$, one can write
$$D(\omega):= \sum_{m=1}^M S_m = \sum_{m=1}^M\angle \tilde{\tau}_m\cdot\left( 1 + \sum_{j=1}^{A_i} \angle \tilde{\varphi}_{mj}\right).$$
The statistical assumptions on $\tilde{\tau}_i$ and $\tilde{\varphi}_{ij}$ imply that $S_1,\dots, S_M$ are an i.i.d.~sample conditional on $M$ (this is not true when conditioning on the $A_i$), therefore $D(\omega)|M$ is a sum of independent samples (the $S_i$).

We collect some properties that the $S_i$ and $D(\omega)$ have, under the null of the Schuster test:

\begin{Lem}\label{Lem:Si}
Under the null assumption of $\angle \tilde{\tau}_i \sim \Unif \{x\in \CC\;:\;|x|=1\}$ (for all $i$), it holds that:
\begin{enumerate}
\itemsep-0.2em
\item[(i)] $S_i$ is circular symmetric, for any $i=1,\dots, M$.
\item[(ii)] $D(\omega)|M = m$ is circular symmetric, for any $m\in \NN$.
\item[(iii)] $D(\omega)$ is circular symmetric.
\end{enumerate}
\end{Lem}
\begin{proof}
(i) this is implied by Proposition~\ref{Prop:circrv}~(ii) and circularity of the unit circle uniform, also see Lemma~\ref{Lem:unifcirc}~(i) and~(ii).\\
(ii) $D(\omega)|M = S_1+\dots S_M$, so $D(\omega|M)$ is circular symmetric by Lemma~\ref{Prop:circrv}~(iii).\\
(iii) Two pairs of distributions are identical iff all conditionals are identical, therefore (applying this to copies of $D(\omega)$ where $M$ remains the same), the fact that $m$ is arbitrary in (ii) implies circular symmetry of $D(\omega)$.
\end{proof}

\subsection{Angle representation of squared Schuster distance}
\label{sec:app-proof.SchusterD}

The squared Schuster distance is defined as $D^2(\omega):= |D(\omega)|^2 = D(\omega)D(\omega)^\ast$. By elementary computation (distributive law and linearity of complex conjugate) we observe that
\begin{align*}
D^2(\omega) & = \sum_{i,j=1}^N \angle \left(\tilde{T}_i-\tilde{T}_j\right)\\
 &= \sum_{m,m'=1}^M \angle \left(\tilde{\tau}_m-\tilde{\tau}_{m'}\right) + 2\Real \sum_{j=1}^{A_m} \angle \left(\tilde{\varsigma}_{mj}-\tilde{\tau}_{m'}\right) +\sum_{j=1}^{A_m}\sum_{j'=1}^{A_{m'}} \angle \left(\tilde{\varsigma}_{mj}-\tilde{\varsigma}_{m'j'}\right)\\
 &= \sum_{m,m'=1}^M \angle \left(\tilde{\tau}_m-\tilde{\tau}_{m'}\right) + 2\Real \sum_{j=1}^{A_m} \angle \left(\tilde{\varsigma}_{mj}-\tilde{\tau}_{m'}\right) +\sum_{j=1}^{A_m}\sum_{j'=1}^{A_{m'}} \angle \left(\tilde{\varsigma}_{mj}-\tilde{\varsigma}_{m'j'}\right)\\
 &= \sum_{\substack{m,m'=1\\ m\neq m'}}^M \angle \tilde{\tau}_m \cdot \left(\angle \tilde{\tau}_{m'}\right)^\ast + 2\Real \sum_{j=1}^{A_m} \angle \tilde{\varsigma}_{mj}\cdot\angle\left(\tilde{\tau}_{m'}\right)^\ast +\sum_{j=1}^{A_m}\sum_{j'=1}^{A_{m'}} \angle \tilde{\varphi}_{mj}\cdot \left(\angle \tilde{\varphi}_{m'j'}\right)^\ast\cdot \angle\tilde{\tau}_m \cdot \left(\angle \tilde{\tau}_{m'}\right)^\ast\\
 & + N + 2 \Real  \sum_{m=1}^M\sum_{j=1}^{A_m} \angle \tilde{\varphi}_{mj}
 +\sum_{\substack{j,j'=1\\ j\neq j'}}^{A_m} \angle \tilde{\varphi}_{mj} \cdot \left(\angle\tilde{\varphi}_{mj'}\right)^\ast,
\end{align*}
where for the last equality we have split double summation based on whether indices are equal or not, and that $\angle 0 = 1$ (which happens in differences if some or all occurring indices are equal).

By the assumptions in Section~\ref{sec:app-proof.setting}, for each sum appearing in the last line, all product factors in the summands are statistically independent of each other.

\subsection{Expectation of squared Schuster distance}
\label{sec:app-proof.Schuster_expectation}
We now proceed with computing the conditional expectation $\EE[D^2(\omega)| M, A_1,\dots, A_m]$ under the null hypothesis assumption of no seasonality at frequency $\omega$, i.e., $\angle \tilde{\tau}_i \sim \Unif \{x\in \CC\;:\;|x|=1\}$.

Under this null hypothesis, recalling that in every sum, all product factors are statistically independent, all sums with a term $\angle \tilde{\tau}_i$ vanish after taking expectations. For example,
$\EE \left[\angle \tilde{\tau}_m \cdot \left(\angle\tilde{\tau}_{m'}\right)^\ast\right] = \EE \left[\angle\tilde{\tau}_m\right] \cdot \EE\left[\angle\tilde{\tau}_{m'}^\ast\right] = 0$, with first equality from independence of $\tilde{\tau}_m$ and $\tilde{\tau}_{m'}$, and second equality from the null assumption. Overall,
\begin{align*}
\EE[D^2(\omega)| M, A_1,\dots, A_m] &= N + 2 \Real  \sum_{m=1}^M\sum_{j=1}^{A_m} \EE[\angle \tilde{\varphi}_{mj}]
 +\sum_{\substack{j,j'=1\\ j\neq j'}}^{A_m} \EE[\angle \tilde{\varphi}_{mj}] \cdot \EE\left[\angle\tilde{\varphi}_{mj'}\right]^\ast\\
  &= N + 2(N-M) \Real \chi_{\varphi}(\omega) + \sum_{m=1}^M A_m(A_m-1) |\chi_{\varphi}(\omega)|^2\\
  &= (N-M) \left(1 - \left|\chi_{\varphi}(\omega)\right|^2\right) + \sum_{m=1}^{M} \left|1 + A_m\cdot\chi_{\varphi}(\omega)\right|^2
\end{align*}
where the second line follows from Lemma~\ref{Lem:Fourier}~(i), and the last line from using that $N-M=\sum_{m=1}^M A_m$.\\

For the unconditional expectation, we use the law of iterated expectation:
\begin{align*}
\EE[D^2(\omega)] &= \EE\left[\EE[D^2(\omega)| M, A_1,\dots, A_m]\right]\\
  &= \EE\left[N + 2(N-M) \Real \chi_{\varphi}(\omega) + \sum_{m=1}^M A_m(A_m-1) |\chi_{\varphi}(\omega)|^2\right]\\
  &= \EE\left[\EE\left[N + 2(N-M) \Real \chi_{\varphi}(\omega) + \sum_{m=1}^M A_m(A_m-1) |\chi_{\varphi}(\omega)|^2\middle|M\right]\right]\\
  &= \EE\left[M+M\nu + 2 M\nu\cdot \Real \chi_{\varphi}(\omega) + \sum_{m=1}^M \nu^2 |\chi_{\varphi}(\omega)|^2 \right]\\
  &= \nu_0+\nu_0\nu + 2 \nu_0\nu\cdot \Real \chi_{\varphi}(\omega) + \nu_0\nu^2 |\chi_{\varphi}(\omega)|^2 \\
  &= \nu_0\nu + \nu_0 \cdot \left| 1 + \nu\cdot \chi_{\varphi}(\omega)\right|^2
\end{align*}
where we have repeatedly used the law of iterated expectation, properties of the Poisson random variables $M\sim \Poi(\nu_0)$ and $A_m|M\sim \Poi(\nu)$, which were also be used as identities $\EE[N|M] = M + M\nu,$ and $\EE[A_m(A_m-1)|M] = \nu^2,$ to obtain the fourth line from the third.

We summarize the above in a proposition

\begin{Prop}\label{prop:exd-appx}
Assume that there is no periodicity at frequency $\omega$, that is, $\angle \tilde{\tau}_i \sim \Unif \{x\in \CC\;:\;|x|=1\}$ for all $i$. Then:
\begin{enumerate}
\itemsep-0.2em
\item[(i.a)] it holds for the expected Schuster distance that $\EE[D(\omega)] = 0$
\item[(i.b)] it holds for the variance of the absolute Schuster distance that\\
$\Var[|D(\omega)|] = \Var[D(\omega)] = \EE[D^2(\omega)]$
\item[(ii)] it holds for the expected squared Schuster distance spectrum that
\begin{align}\label{eq:prop1Appendix}
	\EE[D^2(\omega)] &= \nu_0\cdot\nu + \nu_0\cdot\left|1+ (\calF\lambda)\left(\frac{\omega}{2\pi}\right)\right|^2 \nonumber\\
	&= \nu_0\cdot\nu + \nu_0\cdot\left|1+ (\calF\lambda)\left(k^{-1}\right)\right|^2,
\end{align}
where $\calF\lambda$ denotes the Fourier transform (unitary normalization convention) of the aftershock intensity functional $\lambda$, and $\nu$ is the rate constant of the aftershock process, i.e., $\nu =\int_0^\infty \lambda(t)\diff t.$
\item[(iii)] if the numbers $M$ of primary shocks and $A_1,\dots, A_M$ of aftershocks are considered fixed, the expected squared Schuster distance (conditional on observed numbers) can be expressed as
\begin{align}
	\EE[D^2(\omega)|M,A_1,\dots, A_M] = (N-M) \left(1 - \left|\hat{g}\left(k^{-1}\right)\right|^2\right) + \sum_{m=1}^{M} \left|1 + A_m\cdot \hat{g}\left(k^{-1}\right)\right|^2,
\end{align}
where $\hat{g}:=\nu^{-1} (\calF \lambda),$ i.e., $\hat{g}$ is the Fourier transform of the pdf $\nu^{-1}\lambda$ (i.e., the function $t\mapsto \nu^{-1}\lambda(t),$ which is the canonical aftershock pdf according to Theorem~\ref{Thm:osp}).
\item[(iv)] if the numbers $M$ of primary shocks is considered fixed, the expected squared Schuster distance (conditional on observed numbers) can be expressed as
\begin{align}
	\EE[D^2(\omega)|M] = M\nu + M \left| 1 + (\calF\lambda)\left(k^{-1}\right)\right|^2 .
\end{align}
\end{enumerate}
\end{Prop}
\begin{proof}
(i.a) This is implied by Lemma~\ref{Lem:Si}~(iii) and Proposition~\ref{Prop:circrv}~(iv).
(i.b) This is almost immediate from (i.a) and definition of complex variance; or, more explicitly, from (i.a), and equivalence of (i) and (ii) in Lemma~\ref{Lem:complexvar}.\\
(ii)-(iv) are proved in the in-text calculation above.
\end{proof}

\subsection{The aftershock corrected Schuster test}
\label{sec:app-schuster-test-aft}

To derive an aftershock corrected Schuster test, we need to derive a test statistic.

Writing $S_i:= \angle \tilde{\tau}_i\cdot\left( 1 + \sum_{j=1}^{A_i}+ \angle \tilde{\varphi}_{ij}\right)$, as in Section~\ref{sec:app-proof.SchusterDsymbols}, one can write
$D(\omega)= \sum_{m=1}^M S_m $ as an sum of the i.i.d.~sample $S_1,\dots, S_M$, conditional on $M$.
The pair of real random variables $\Real D(\omega)|M$ and $\Imag D(\omega)|M,$ is hence subject to well-known results on sums of independent samples, and we can pursue the same strategy as in the original Schuster test, see Section~\ref{sec:app-schuster.distasy}, with $M$ governing the asymptotic instead of $N$.

For this, we define $Z_i:=(\Real S_i, \Imag S_i), 1\le i\le M$. For ease of notation, we also introduce generative versions $S, Z$ such that $Z_1,\dots, Z_M\sim Z$ and $S_1,\dots, S_M\sim S$. We need to compute expectation and variance of $Z_i$ to use the central limit theorem:

\begin{Lem}\label{Lem:covrealD}
For the bivariate real random variable $Z$, it holds under the null of $\angle \tilde{\tau}_i \sim \Unif \{x\in \CC\;:\;|x|=1\}$ (for all $i$) that:
\begin{enumerate}
\itemsep-0.2em
\item[(i)] $\EE[S] = 0$, and $\EE [Z] = (0,0)$
\item[(ii)] $\EE[D(\omega)|M] = 0$
\item[(iii)] $\Var[D^2(\omega)|M] = \EE[D^2(\omega)|M]$
\item[(iv)] $\Var[S] = \EE[D^2(\omega)|M]/M = \nu + \left| 1 + (\calF\lambda)\left(k^{-1}\right)\right|^2.$
\item[(v)] $\Var [Z] = \frac{\EE[D^2(\omega)|M]}{2M}\cdot I = \nu \cdot I + \left| 1 + (\calF\lambda)\left(k^{-1}\right)\right|^2\cdot I,$ where $I\in \RR^{2\times 2}$ is the bivariate identity matrix.
\end{enumerate}
\end{Lem}
\begin{proof}
(i) This is implied by Lemma~\ref{Lem:Si}~(iii) and Proposition~\ref{Prop:circrv}~(i), observing that $S$ and $S_1$ are identically distributed.\\
The statement for $Z$ then follows from observing that this implies $\EE[ Z_1] = \EE [(\Real S_1, \Imag S_1)] = (0,0)$, which implies $\EE[Z] = 0$ as $Z$ and $Z_1$ are identically distributed.\\
(ii) This follows from (i) and linearity of expectation.\\
(iii) This is due to (ii) and equivalence of (i) and (ii) in Lemma~\ref{Lem:complexvar}.\\
(iv) By Lemma~\ref{Lem:complexvarprops}~(ii), $\Var [S]\cdot M = \Var[D(\omega)|M]$. Also, , proving the first equality. The second equality follows from Proposition~\ref{prop:exd-appx}~(iv).\\
(v) The second equality follows from (ii); the first equality follows from Lemma~\ref{Lem:Si}~(ii) and Proposition~\ref{Prop:circrv}~(v).
\end{proof}

As a direct corollary of Lemma~\ref{Lem:covrealD} and the central limit theorem, we obtain:

\begin{Cor}
Define $\overline{Z}_M:= (\Real D(\omega),\Imag D(\omega))/M.$
Under the null hypothesis of $\angle \tilde{\tau}_i \sim \Unif \{x\in \CC\;:\;|x|=1\}$ (for all $i$), it holds that:
\begin{enumerate}
\itemsep-0.2em
\item[(i)] $\EE[\overline{Z}_M|M] = 0$, and $\Var[\overline{Z}|M] = \Var [Z]/M = \frac{\nu}{M} + \frac{1}{M}\left| 1 + (\calF\lambda)\left(k^{-1}\right)\right|^2$.
\item[(ii)] $\sqrt{M}\cdot\overline{Z}_M|M=m \overset{d}{\rightarrow} \calN(0,\Var [Z])$ as $m\rightarrow \infty$
\end{enumerate}
\end{Cor}

In particular, under the null, and writing $X:= \Real D(\omega),\; Y:=\Imag D(\omega)$, we can infer that $(X/\sqrt{M}, Y/\sqrt{M}\approx \calN(0, \frac{1}{2}\Var [Z])$ is a good approximation as long as the actual value of $M$ is large (say, $\ge 30$). We note two key differences to the Schuster test not in the presence of aftershocks as described in Section~\ref{sec:app-schuster.distasy}: the asymptotic is in the number of primary shocks $M$, not in the number of shocks overall; the covariance matrix in the approximation is still isotropic, but has value $\Var [Z] = \frac{\nu}{2} \cdot I + \frac{1}{2}\left| 1 + (\calF\lambda)\left(k^{-1}\right)\right|^2\cdot I,$ which in general is different from the aftershock-free value $\frac{1}{2} I$.\\

Normalizing instead with the easily estimable (by $N$) expected number of total shocks, $M(\nu+1)$, we obtain that
$$(X, Y)/\sqrt{M(\nu+1)}\approx \calN(0, \frac{1}{2(\nu+1)}\Var [Z]) = \calN\left(0,\; \frac{1}{2} \cdot I-\frac{1}{2\kappa}\cdot I + \frac{1}{2 \kappa}\left| 1 + (\calF\lambda)\left(k^{-1}\right)\right|^2\cdot I\right),$$
where we have written $\kappa := \nu + 1$, the expected number of shocks per primary shock.
This exposes two isotropic additive corrections to the aftershock-free variance with opposite sign: a negative term with scale $\kappa^{-1}$, and a positive term with scale $\kappa^{-1}\left| 1 + (\calF\lambda)\left(k^{-1}\right)\right|^2$.

In either case, $\frac{1}{M}D^2(\omega)|M \approx \Exp\left(\nu + \left| 1 + (\calF\lambda)\left(k^{-1}\right)\right|^2\right),$ according to Lemma~\ref{Lem:normexp} (and continuous mapping theorems), which exhibits a smooth dependence of the Schuster spectrum on the Fourier spectrum of $\lambda$.

\end{document}